\documentclass[11pt,a4paper]{article}
\pdfoutput=1
\usepackage{graphicx}

\usepackage{amsmath}
\usepackage{amssymb,amsfonts}
\usepackage{lineno}
\usepackage[mathscr]{eucal}
\usepackage{eucal}
\usepackage{subfigure}
\usepackage{float}
\def\nevts{65}

\title{Results of a self-triggered prototype system for radio-detection of extensive air showers at the Pierre Auger Observatory}

\begin{document}
\author{The Pierre Auger Collaboration\footnote{Authors are listed on the
following pages.}, S. Acounis$^{31}$, D. Charrier$^{31}$, T. Gar\c con$^{31}$, C. Rivi\`ere$^{29}$, P. Stassi$^{29}$.}
\maketitle

\abstract
{We describe the experimental setup and the results of RAuger, a small radio-antenna array, consisting of three fully autonomous and self-triggered radio-detection stations, installed close to the center of the Surface Detector (SD) of the Pierre Auger Observatory in Argentina. The setup has been designed for the detection of the electric field strength of air showers initiated by ultra-high energy cosmic rays, without using an auxiliary trigger from another detection system. Installed in December 2006, RAuger was terminated in May 2010 after \nevts\ registered coincidences with the SD. The sky map in local angular coordinates (i.e., zenith and azimuth angles) of these events reveals a strong azimuthal asymmetry which is in agreement with a mechanism dominated by a geomagnetic emission process. The correlation between the electric field and the energy of the primary cosmic ray is presented for the first time, in an energy range covering two orders of magnitude between 0.1~EeV and 10~EeV. It is demonstrated that this setup is relatively more sensitive to inclined showers, with respect to the SD.
In addition to these results, which underline the potential of the radio-detection technique, important information about the general behavior of self-triggering radio-detection systems has been obtained. In particular, we will discuss radio self-triggering under varying local electric-field conditions.}

\newpage
\par\noindent
{\bf The Pierre Auger Collaboration} \\
P.~Abreu$^{61}$, 
M.~Aglietta$^{49}$, 
M.~Ahlers$^{91}$, 
E.J.~Ahn$^{79}$, 
I.F.M.~Albuquerque$^{15}$, 
I.~Allekotte$^{1}$, 
J.~Allen$^{83}$, 
P.~Allison$^{85}$, 
A.~Almela$^{11,\: 7}$, 
J.~Alvarez Castillo$^{54}$
% ...

\par\noindent
$^{1}$ Centro At\'{o}mico Bariloche and Instituto Balseiro (CNEA-UNCuyo-CONICET), San 
Carlos de Bariloche, 
Argentina \\
% ...
$^{7}$ Instituto de Tecnolog\'{\i}as en Detecci\'{o}n y Astropart\'{\i}culas (CNEA, CONICET, UNSAM), 
Buenos Aires, 
Argentina \\
% ...
$^{11}$ Universidad Tecnol\'{o}gica Nacional - Facultad Regional Buenos Aires, Buenos Aires,
Argentina \\
% ...
$^{15}$ Universidade de S\~{a}o Paulo, Instituto de F\'{\i}sica, S\~{a}o Paulo, SP, 
Brazil \\
% ...
$^{49}$ Istituto di Fisica dello Spazio Interplanetario (INAF), Universit\`{a} di Torino and 
Sezione INFN, Torino, 
Italy \\
% ...
$^{54}$ Universidad Nacional Autonoma de Mexico, Mexico, D.F., 
Mexico \\
% ...
$^{61}$ LIP and Instituto Superior T\'{e}cnico, Technical University of Lisbon, 
Portugal \\
% ...
$^{79}$ Fermilab, Batavia, IL, 
USA \\
% ...
$^{83}$ New York University, New York, NY, 
USA \\
% ...
$^{85}$ Ohio State University, Columbus, OH, 
USA \\
% ...
$^{91}$ University of Wisconsin, Madison, WI, 
USA \\
% ...
\par\noindent
(\ddag) Deceased \\
(a) Now at Konan University \\
(b) Also at the Universidad Autonoma de Chiapas on leave of absence from Cinvestav \\
(c) Now at University of Maryland \\
(d) Now at NYU Abu Dhabi \\

\newpage

\section{Introduction}
The origin of cosmic rays at ultra high energies ($E > 0.1$~EeV) is a fundamental question of astroparticle physics. The induced shower of secondary particles in the atmosphere of the Earth provides essential information on the cosmic ray itself: arrival direction, primary energy, and mass. An established method to assess the mass of the primary particle is based on the determination of the longitudinal shower profile, for instance by using optical detectors which collect the ultra-violet photons emitted by excited nitrogen along the path of the shower (see, e.g., \cite{AugerFDcompo}). As this light signal is very weak, it can only be observed during dark nights, limiting the duty cycle for this detection technique to about 14\%.
In the air shower many electrons and positrons are created, forming a pancake-shaped particle front with a typical thickness ranging from less than 1~m close to the shower axis to more than 10~m far from the shower axis.
The geomagnetic field induces a drift velocity in these particles which is perpendicular to the direction of the initial cosmic ray and which is in opposite directions for electrons or positrons. The strength of this current is roughly proportional to the number of charged particles. As this number changes while the shower develops through the atmosphere, coherent emission of electromagnetic waves occurs at wavelengths larger than the size of the dimension of the charge cloud, i.e., for radio frequencies less than about 300~MHz, which are in the VHF band. Thus, while the fluorescence light is proportional to the energy deposit,
radio signals probe the increase and the decrease of the number of electrons and positrons in the shower. Therefore, radio signals carry information which is complementary to that from the observation by fluorescence emission, as well as to that from the detection of secondary particles hitting the surface of the Earth \cite{huegesim,mgmr}. Since radio waves are hardly affected by their passage through the atmosphere, a radio-detection array has a potential duty cycle of almost 100\%, although effects of atmospheric disturbances (thunderstorms) \cite{lopesICRC2009} and transient electromagnetic interferences (human-related activities) reduce this quantity. In the last two years, substantial progress has been made in radio detection of cosmic rays, technically through the Auger Engineering Radio Array (AERA) project~\cite{aera10} which benefits from the results obtained by the CODALEMA~\cite{coda05,coda09} and the LOPES~\cite{lopes05,lopes10} experiments. The AERA project was preceded by important tests performed with small-scale experiments to detect cosmic rays at the Pierre Auger Observatory and which were used to further develop the radio-detection technique for large-scale experiments.

This report describes the RAuger setup which ran in its first version between December 2006 and May 2010.
RAuger uses the same antennas as the ones used in CODALEMA but differs in all other aspects: RAuger is fully autonomous and independent of any external detector, contrary to CODALEMA where the radio array is triggered by a particle detector and powered by cables.
We will use the angular coordinates $\theta$ and $\phi$ as the zenith and the azimuth angles, respectively, and where $\phi = 0^\circ$ $(90^\circ)$ denotes the geographic east (north). In section~\ref{expsetup}, we describe the RAuger experimental setup, and in section~\ref{electricFieldDependence} the dependence of the trigger rate on the local electric field conditions, and we discuss the possible consequences for event selection. Finally, in section~\ref{augerCoincidences}, the events detected by the Surface Detector (SD) of the observatory which were in coincidence with the events registered by the radio prototype are presented.
Initial data obtained with this setup have been reported in~\cite{revenu08,revenu10}.
In the past, radio detection of cosmic rays has been discussed in~\cite{hazen,hough,selftriggergransasso1,selftriggergransasso,green}.
More recently, similar reports have been obtained from other experiments as well \cite{ANITA,CHINA,ardouin}.

\section{The RAuger experimental setup: radio-detection prototype stations}\label{expsetup}
\begin{figure}[!ht]
\center
\includegraphics[width=10cm]{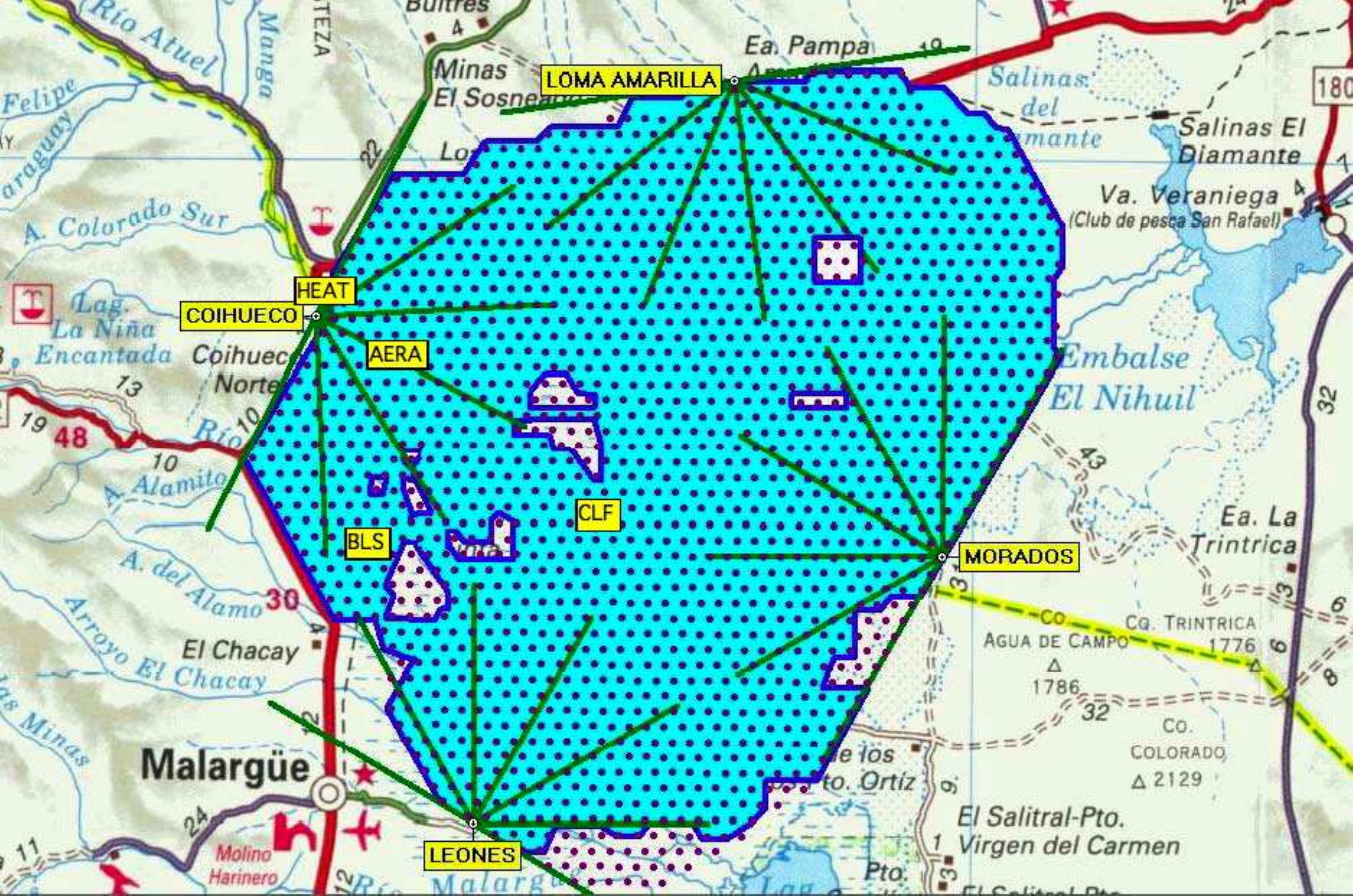}
\caption{\sl{Map of the Pierre Auger Observatory area (see the online colored version), showing the stations of the SD (blue dots), the sites of the telescopes of the FD (Los Leones, Los Morados, Loma Amarilla and Coihueco), the location of the Auger Engineering Radio Array (AERA), and the sites where prototypes of AERA have been deployed: Balloon Launching Station (BLS) and Central Laser Facility (CLF). The distance between Loma Amarilla and Los Leones is 64~km.}\label{fig:augerLayout}}
\end{figure}

The Pierre Auger Observatory is located near Malarg\"ue, in the province of Mendoza in Argentina. In its basic layout, it is a hybrid detector composed of the SD \cite{augersd} and the Fluorescence Detector (FD) \cite{AugerFDcompo}. The SD is composed of 1660 water Cherenkov detectors arranged as an array on a triangular grid with 1.5~km spacing. An elementary triangle has an area of $0.97$~km$^{2}$. In Figure~\ref{fig:augerLayout}, the various components of the observatory are displayed. The SD-determined energy $T_{50}$, where this detector has 50\% detection efficiency, is about 1~EeV.

\begin{figure}[!ht]
\begin{minipage}{20cm}
\begin{center}
\includegraphics[width=6.5cm]{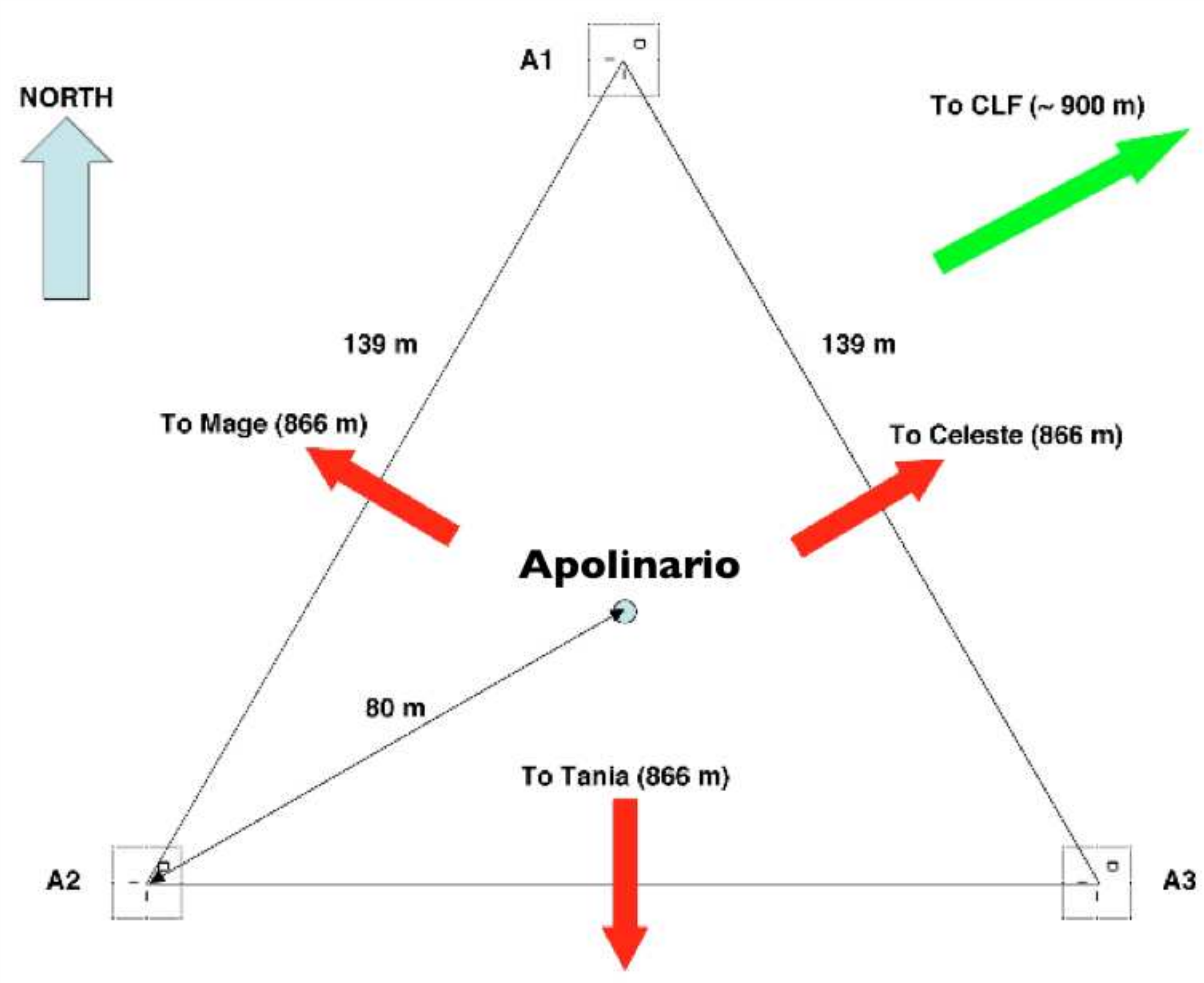}
\hspace{1.5cm}
\includegraphics[width=6cm]{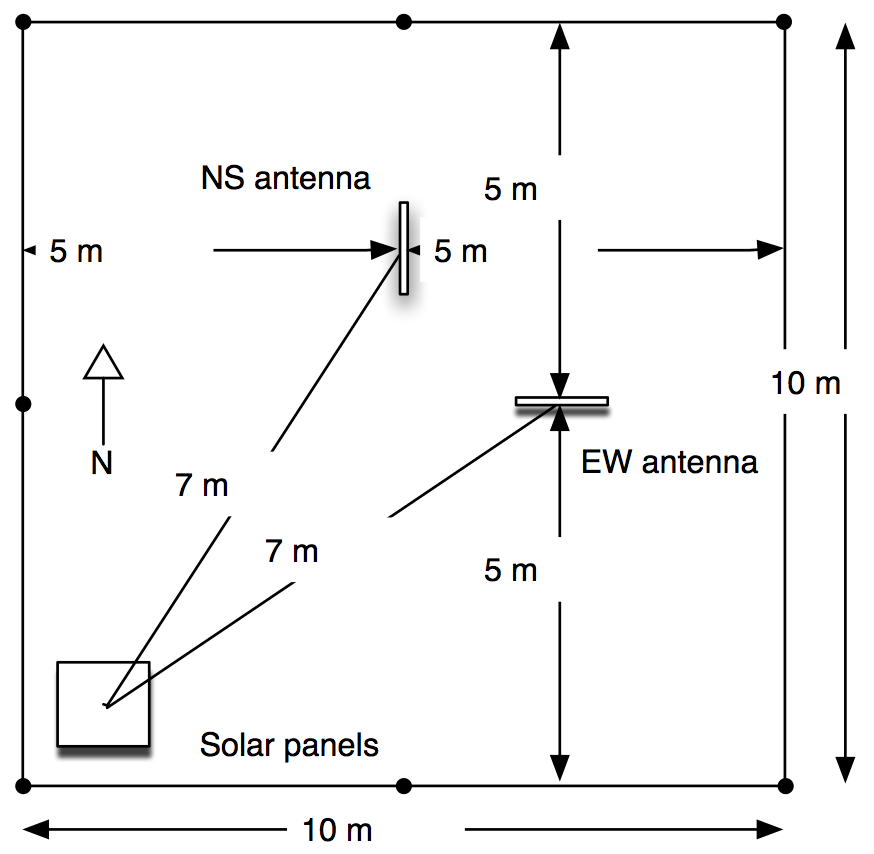}
\end{center}
\end{minipage}
\caption{\sl{Left: setup of the RAuger experiment, with the three radio-detection stations A1, A2 and A3 around the SD station called Apolinario, where stations Mage, Celeste, and Tania are the three neighboring stations of the SD basic layout. Right: sketch of an individual radio-detection station with two dipole antennas, one aligned north-south (NS) and one aligned east-west (EW).}}\label{fig:generalsetup}
\end{figure}

\begin{figure}[!ht]
\center
\includegraphics[width=8cm]{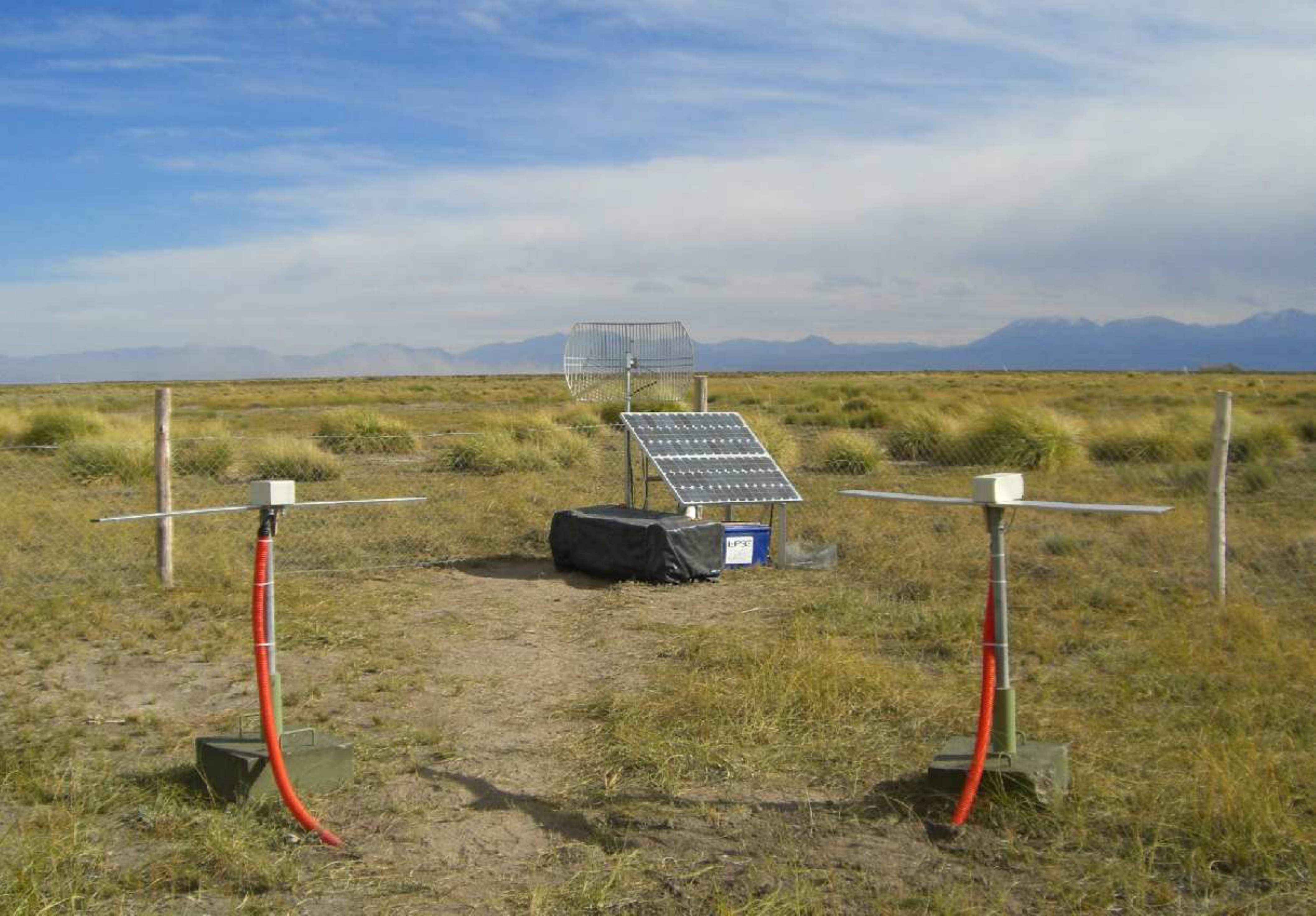}
\caption{\sl{Photo of a radio-detection station with its two dipole antennas and, in the back, the solar panels, the electronics box (covered by a black plastic sheet) and the WiFi antenna pointing toward the CLF.}}\label{fig:stationpicture}
\end{figure}

RAuger \cite{revenu08,revenu10} was composed of three fully autonomous radio-detection stations (A1, A2, A3). Each station was independent in terms of power supply, trigger, data acquisition, and data transmission. The layout of the three radio-detection stations is presented in Figure~\ref{fig:generalsetup}, whereas Figure~\ref{fig:stationpicture} shows a photo of one of these stations. The three stations were placed near the center of the SD array close to the CLF, at the corners of a small equilateral triangle, with an area of 8400~m$^2$ representing 0.86\% of an SD elementary triangle. At the center of the nearest elementary triangle, an additional SD station (named Apolinario) was installed to locally increase the SD event rate. The coincident events involving three SD stations including Apolinario were used for arrival direction studies only, because the full reconstruction of the corresponding shower is not reliable for core position and energy estimation. The coincident events involving three or more regular SD stations are  fully usable (note that the Apolinario data were not used in the reconstruction algorithms in order to avoid parameter biasing with respect to the SD dataset).

\subsection{Antennas}\label{secant}
\begin{figure}[!ht]
\center
\includegraphics[width=14cm]{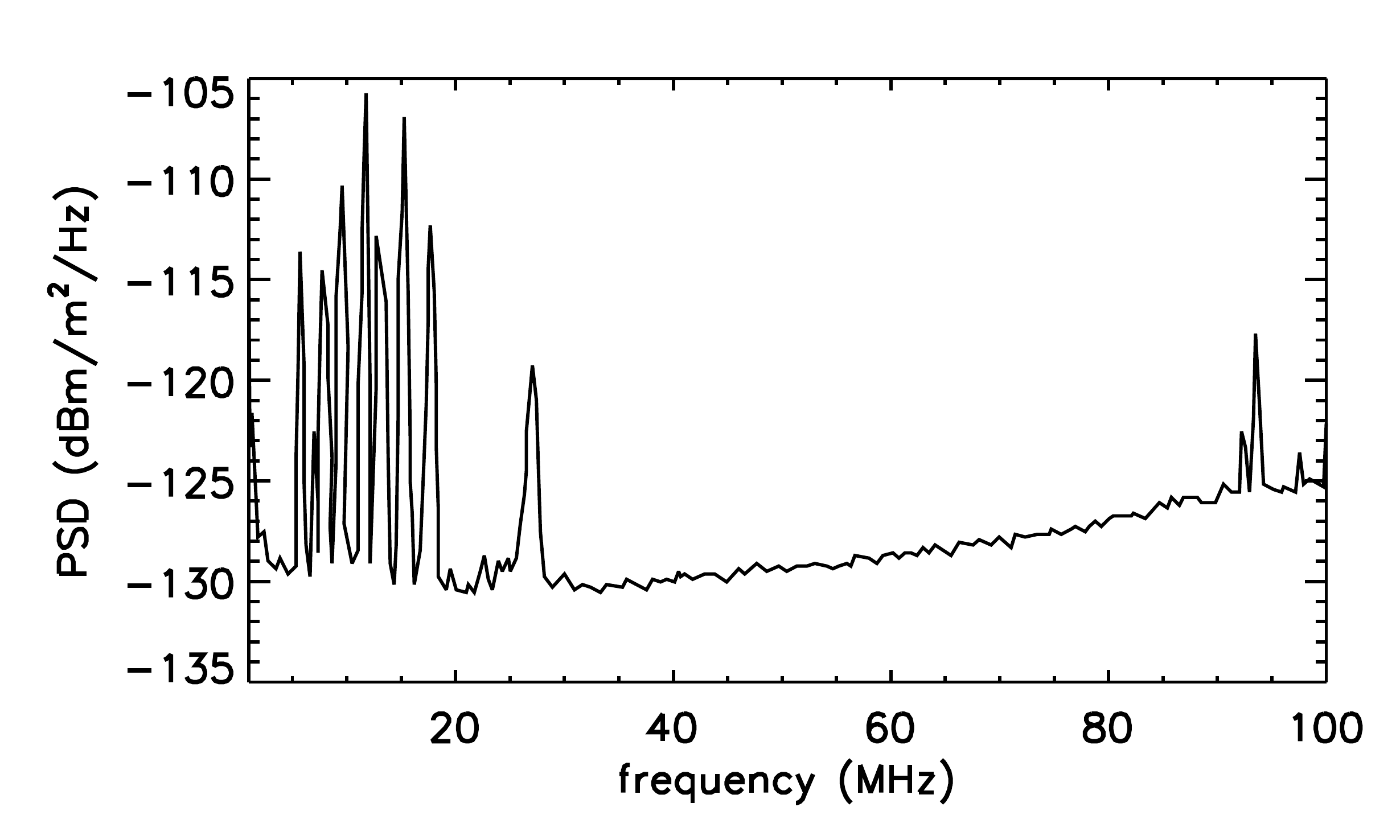}
\caption{\sl{Spectrum measured with the dipole antenna at the same location as the RAuger experiment. The antenna response is not deconvoluted. The strong emission line visible here at $\sim 27$~MHz is not always present in the data. The frequency band 40-80~MHz is, most of the time, free of emitters.}}\label{backsp}
\end{figure}

To detect the very fast electric field-transients produced by air showers, which have a typical pulse width of 10 - 100 ns~\cite{allan}, we used a wide-band antenna made of a dipole receiver coupled to a dedicated front-end amplifier. Therefore, this dipole antenna acted as an active one and not as a simple short dipole. The amplifier was a low-noise, high input impedance, dedicated ASIC with a gain of 34 dB from 0.1~MHz to 200~MHz, with a $3$~dB bandwidth from 0.08~MHz to 230~MHz. Our frequency range of interest was 20-80~MHz, and a low-pass filter was inserted after the front-end amplifier to suppress high frequency and very powerful TV transmitter carriers above 200~MHz, that would otherwise have added a strong noise component to any cosmic-ray signal.
Figure~\ref{backsp} presents a spectrum recorded at the CLF, using background data at the output of the antenna. The antenna response is not deconvoluted because the background is coming from all directions, which explains the overall shape of the spectrum and its base level. The AM and FM emitters are clearly visible below 20~MHz and above 80~MHz, respectively. 
This type of antenna has been in use at the CODALEMA experiment~\cite{dipolar} since 2005. The dipole was made of two aluminium slats of 0.6~m length and 0.1~m width and was installed horizontally at 1~m above ground. Additional and detailed technical information can be found in~\cite{dipolar}. The frequency and directivity response, which includes the amplifier response to describe the ``active antenna" properties, are well understood through both measurements and simulations, which allows us to correct the registered signals for the antenna response. From the measured values $V^{\text{EW}}_m$ and $V^{\text{NS}}_m$ of the voltages in both EW and NS dipoles, we can reconstruct the values of the electric field in the EW and NS directions:
\begin{gather}
E_\text{EW}(t) = \cos\theta\,\cos\phi\,{\cal F}^{-1}\left(
	\frac{H^{\text{NS}}_\phi\,{\cal{F}}(V^{\text{EW}}_m)-H^{\text{EW}}_\phi\,{\cal F}(V^{\text{NS}}_m)}{H^{\text{NS}}_\phi\,H^{\text{EW}}_\theta-H^{\text{EW}}_\phi\,H^{\text{NS}}_\theta}\right)\nonumber\\
+\sin\phi\,{\cal F}^{-1}\left(\frac{H^{\text{NS}}_\theta\,{\cal F}(V^{\text{EW}}_m)-H^{\text{EW}}_\theta\,{\cal F}(V^{\text{NS}}_m)}{H^{\text{NS}}_\phi\,H^{\text{EW}}_\theta-H^{\text{EW}}_\phi\,H^{\text{NS}}_\theta}\right)
\label{eq1}\\
E_\text{NS}(t) = \cos\theta\,\sin\phi\,{\cal F}^{-1}\left(
	\frac{H^\text{NS}_\phi\,{\cal{F}}(V^{\text{EW}}_m)-H^{\text{EW}}_\phi\,{\cal F}(V^{\text{NS}}_m)}{H^{\text{NS}}_\phi\,H^{\text{EW}}_\theta-H^{\text{EW}}_\phi\,H^{\text{NS}}_\theta}\right)\nonumber 
	\\
	-\cos\phi\,{\cal F}^{-1}\left(\frac{H^{\text{NS}}_\theta\,{\cal F}(V^{\text{EW}}_m)-H^{\text{EW}}_\theta\,{\cal F}(V^{\text{NS}}_m)}{H^{\text{NS}}_\phi\,H^{\text{EW}}_\theta-H^{\text{EW}}_\phi\,H^{\text{NS}}_\theta}\right)\label{eq2}
\end{gather}
where $\cal{F}$ is the Fourier transform, $(\theta,\phi)$ the incoming direction and $H^{\text{EW,NS}}_{\theta,\phi}$ are the transfer functions accounting for the antenna and electronics response. $H^{\text{EW,NS}}_{\theta,\phi}$ are complex functions of the frequency and permit one to correct for directivity variation and phase delay. We used the 4NEC2 software~\cite{4nec} to compute these transfer functions for a wide range of arrival directions.

\begin{figure}[!ht]
\begin{minipage}{18cm}
\begin{center}
\includegraphics[width=8cm]{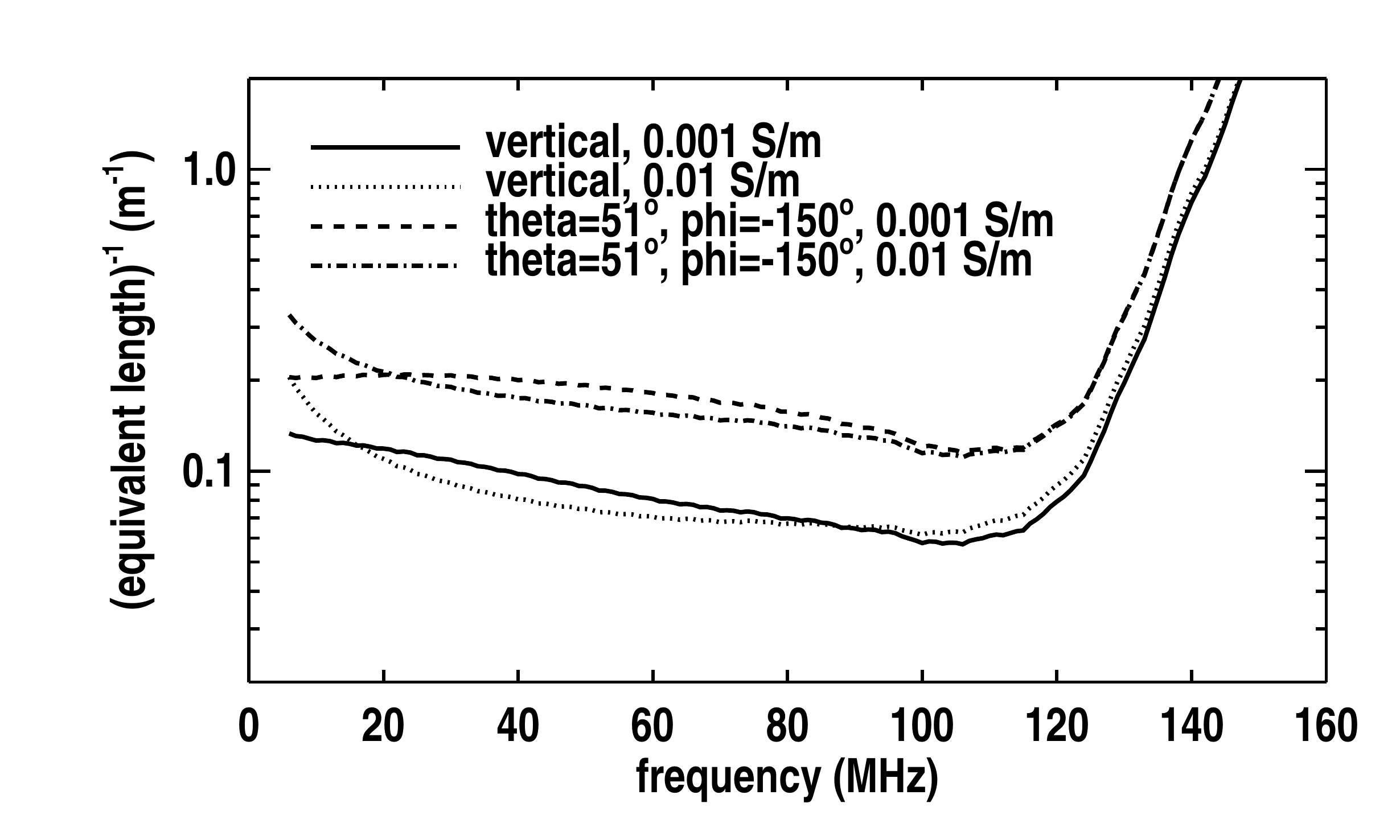}
\includegraphics[width=8cm]{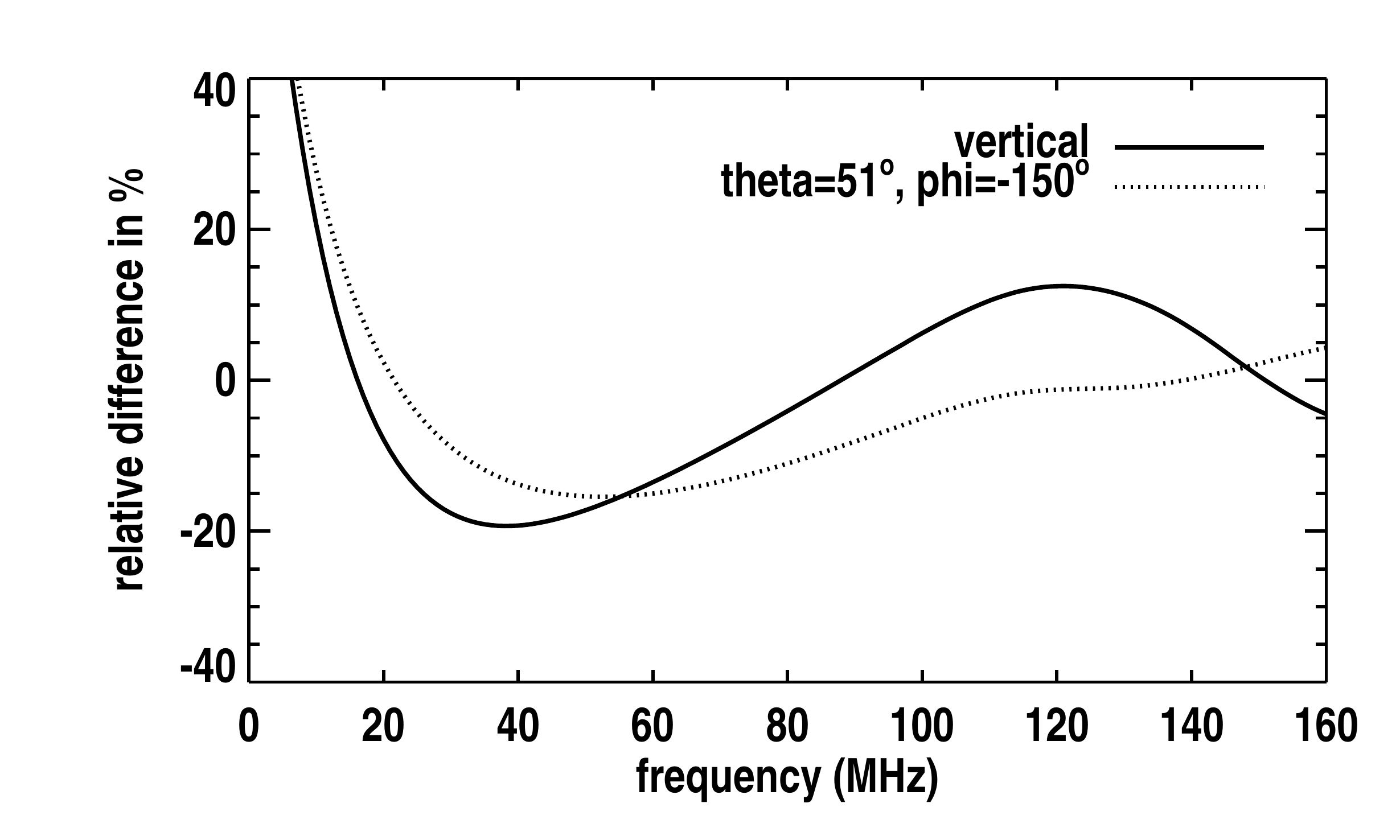}
\includegraphics[width=8cm]{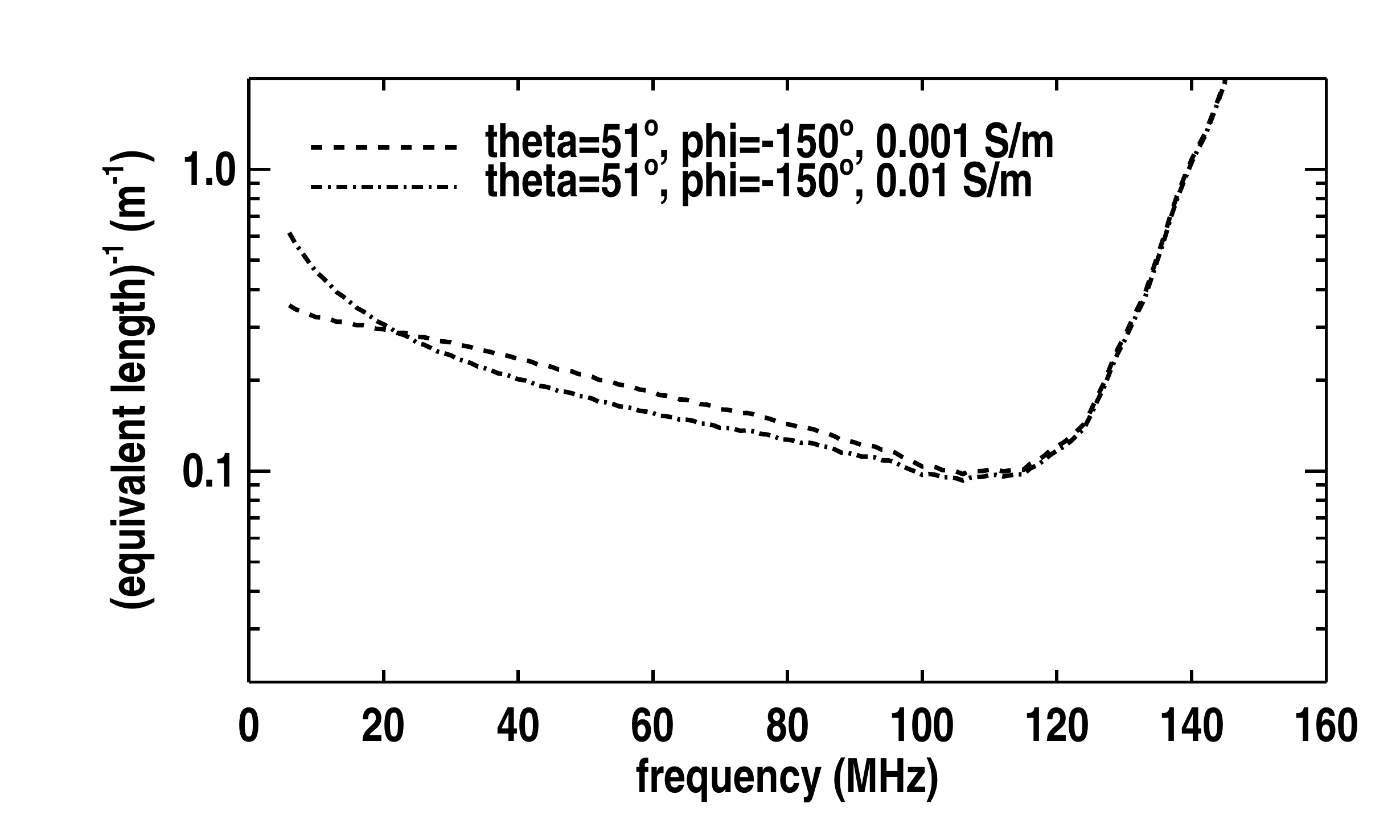}
\includegraphics[width=8cm]{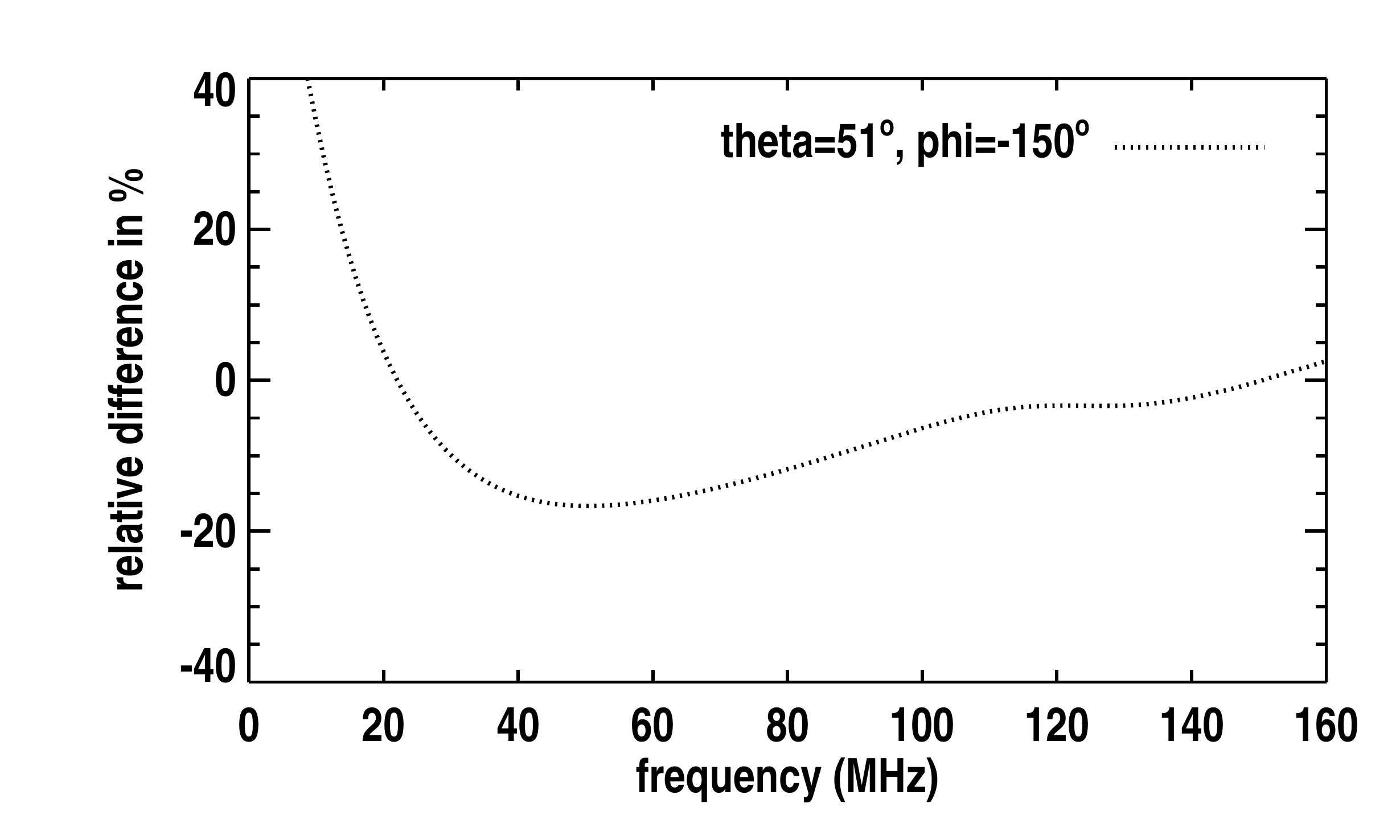}
\end{center}
\end{minipage}
\caption{\sl {Top left: inverse of the equivalent length in the EW direction ($1/|H^{\text{EW}}_\theta|$) as a function of the frequency of a CODALEMA dipole antenna~\cite{dipolar} for two arrival directions and two extreme values $\sigma_\text{min},~\sigma_\text{max}$ of the soil conductivity. Top right: relative difference between the cases $\sigma_\text{min},~\sigma_\text{max}$. Bottom figures: same as top figures for the function $1/|H^{\text{EW}}_\phi|$. The zenith arrival direction corresponds to $H^{\text{EW}}_\phi=0$ and
therefore is not represented in the figure. Our range of frequency of interest is 20-80~MHz.}}
\label{fig:antennaResponse}
\end{figure}

In order to check the effect of the soil properties, we ran simulations with varying values of the conductivity $\sigma$. In 4NEC2 it is not possible to use a conductivity that varies with the frequency. We used instead a value of $\sigma$ constant with the frequency, and we selected two extreme values as reported in~\cite{soil}. We refer to Figures~1 and~2 in this reference corresponding to our case: that of a dry soil. The dry sand dielectric constant can be taken as constant (we used $\varepsilon_r=2.7$, the variations being smaller than 1.8\%) over our frequencies of interest. We cannot use any data below 20~MHz due to AM emitters in the area. The dry sand conductivity varies from $\sigma_\text{min}=9\times 10^{-4}$~S~m$^{-1}$ at 10~MHz up to $\sigma_\text{max}=8\times 10^{-3}$~S~m$^{-1}$ at 80~MHz. For illustration, we computed the transfer functions in the two cases $\sigma_\text{min},~\sigma_\text{max}$ and for two arrival directions (vertical and $\theta=51^\circ,\phi=210^\circ$, corresponding to the threefold event presented in section~\ref{3fold}). Figure~\ref{fig:antennaResponse} (top) displays, as a function of the frequency, the results we obtained for the function $1/|H^{\text{EW}}_\theta|$, which represents the inverse of the equivalent length of the antenna in the EW direction. The results for the function $1/|H^{\text{EW}}_\phi|$ are presented in Figure~\ref{fig:antennaResponse} (bottom). The relative difference between the extreme cases $\sigma_\text{min}$ and $\sigma_\text{max}$ results in a relative error in the inverse equivalent lengths below 20\% for frequencies above 20~MHz. Since we do not have a constant monitoring of the soil properties, we will choose the case $\sigma=\sigma_\text{min}$ for the present analysis and we account for the varying $\sigma$ by considering an uncertainty of 20\% in the transfer function.

\subsection{Electronics and data acquisition}
In addition to the active dipole antennas, the electronics, triggering, and data-acquisition systems used in RAuger contained the following elements:
\begin{itemize}
\item A trigger board with a tunable radio-frequency filter to get rid of frequencies due to human activities;
\item Two channels of a Tektronix THS730A hand-held oscilloscope consisting of an 8-bit Analog to Digital Converter (ADC) adapted to wide-band waveform analysis, working at 500~MS/s for a 5~$\mu$s registered waveform over 2500~points;
\item A GPS Motorola Oncore UT+ receiver (the same as used in the SD stations) for a time reference \cite{auger04};
\item Two 85~W solar panels and two 100~Ah, 12~V batteries for the power supply of the station, while the continuous load for the station electronics was 18~W;
\item A local data-acquisition system based on the Unified Board (UB) developed for the SD stations \cite{auger04}; this UB used the same time tagging system as the SD stations and mastered the local data streams and managed the communication with the remote Radio Data Acquisition System (RDAS) located at the CLF about 900~m east of the radio-detection array;
\item A standard WiFi system (115~kbps) to send the station data to the RDAS.
\end{itemize}

At each station, radio events were recorded by two channels (one per antenna): the first one being the full-band EW signal  between 0.1~MHz and  100~MHz; the second one the full-band NS signal (same frequency boundaries). The trigger decision was made on the EW signal after filtering it between 50 and 70~MHz. This filtered signal was sent to a voltage comparator to build the trigger. One of the important limitations of these stations was that the trigger level could not be changed, neither remotely nor by software; adjustments had to be made by hand.
The acquisition was vetoed until the event was read from the oscilloscope. Subsequently, the data were transmitted via a high-gain WiFi link to the RDAS located in the CLF. Because of a required time of about 2.7~s to read out the oscilloscope and to send these data to the RDAS, the maximum event rate was around 0.37~event~s$^{-1}$. The RDAS received the data from the three stations and no higher-level trigger was used; each registered event was stored on disk and the analysis of the three data streams was done off-line.

\section{Event rate}\label{electricFieldDependence}

For each individual station, the trigger level was adjusted to avoid, as much as possible, saturation of the acquisition rate caused by ambient noise transients. The amplitudes of these transients were rather high even when the local rms noise on the stations was low.

\subsection{Daily cycle}

\begin{figure}[!ht]
\begin{center}
\includegraphics[width=7cm]{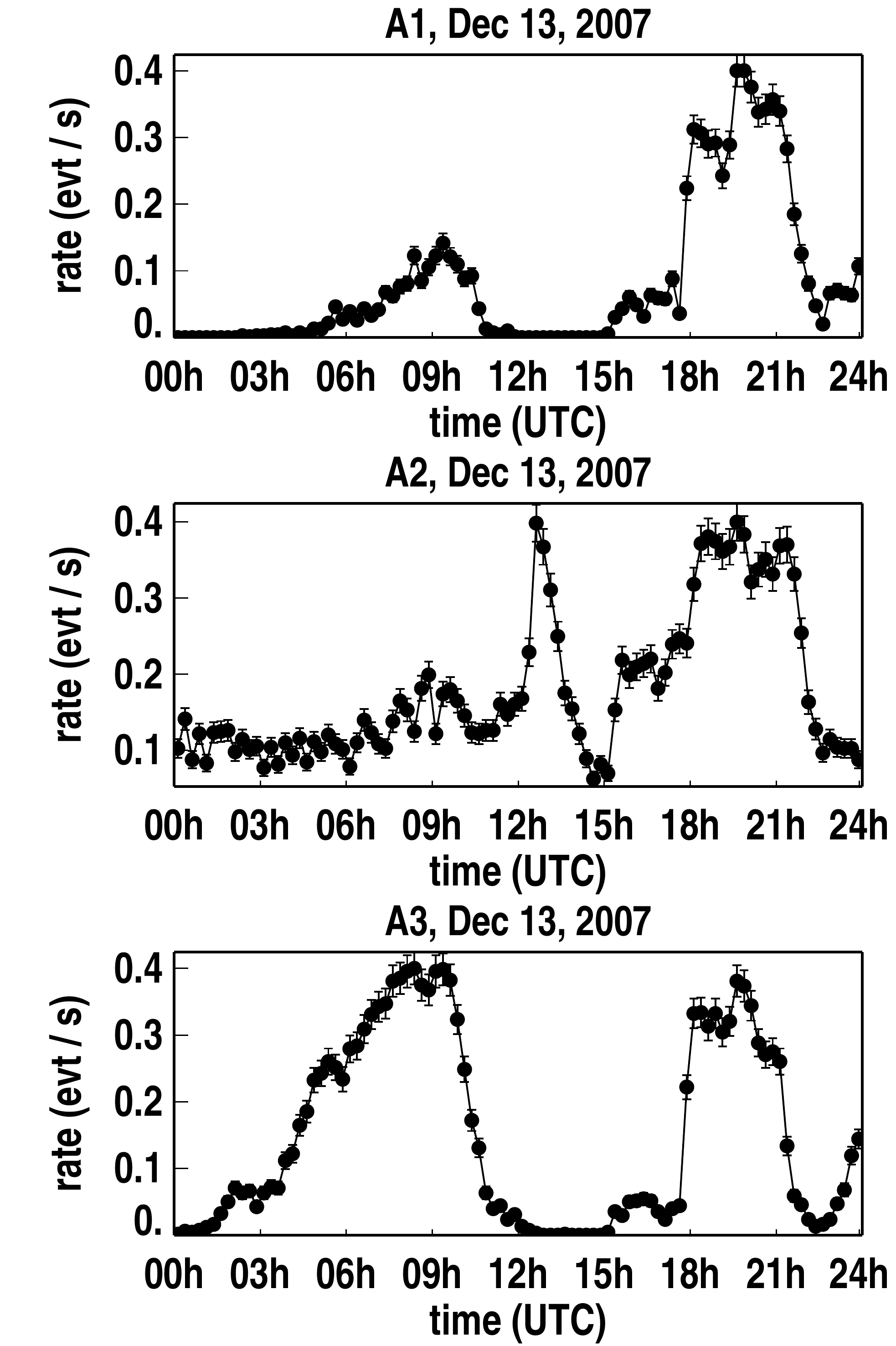}
\includegraphics[width=7cm]{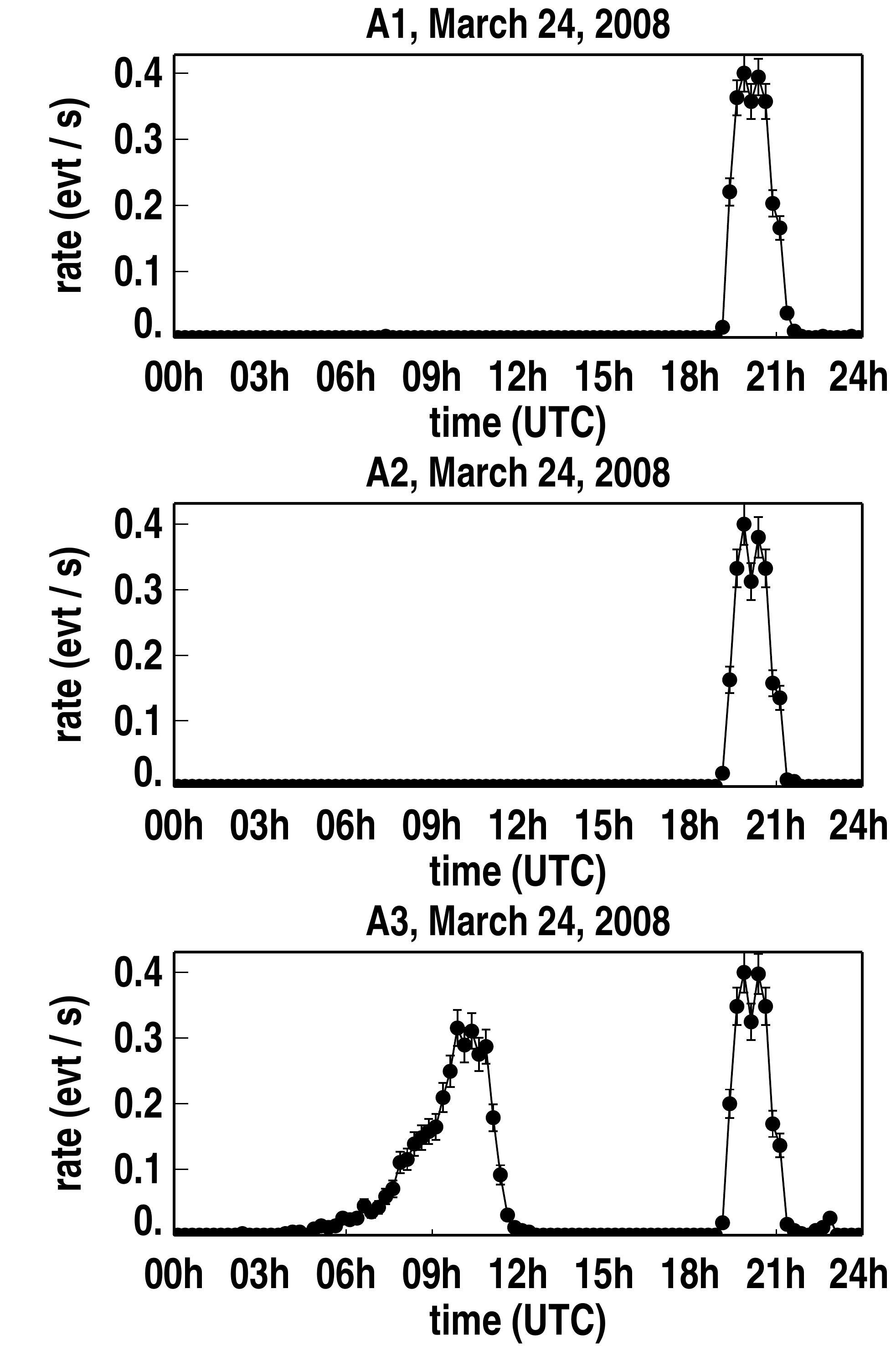}
\end{center}
\caption{\sl{The event rate for the stations A1, A2, and A3 on December 13, 2007 (left) and March 24, 2008 (right). The error bars correspond to the statistical fluctuations in a bin. The horizontal axis is the UTC time; local time in Malarg\"ue, Argentina is UTC-3.}}
\label{fig:trig1}
\end{figure}

In Figure~\ref{fig:trig1}, the behavior of the event rate for the three stations for two typical and very illustrative days is displayed: December 13, 2007 (left panels) and March 24, 2008 (right panels). Between these two days, the thresholds of the three stations were changed effectively by a few $\mu$V~m$^{-1}$,
leading to a significant decrease of the event rate before 15:00~h UTC.
We note that the maximum event rate for the three stations was about 0.37~s$^{-1}$.
A strong sensitivity to noise variations can be seen in Figure~\ref{fig:trig1}.
Furthermore, for a very small change of the detection threshold, one can find triggering conditions leading to a large suppression of the number of events caused by ambient transient noise. It is worth noting that the event rates of the three stations have similar behaviors, but for A3 an additional period of high event rate is visible in Figure~\ref{fig:trig1} during the morning of the two days. It was found that the strength of this bump is correlated with the measured humidity fraction in the air at the CLF, and so it could indicate that A3 was more sensitive to humidity fraction than A1 or A2. A possible cause of this might be different insulation conditions for A3 when the radio stations were installed.

\begin{figure}[!ht]
\begin{center}
\includegraphics[width=10cm]{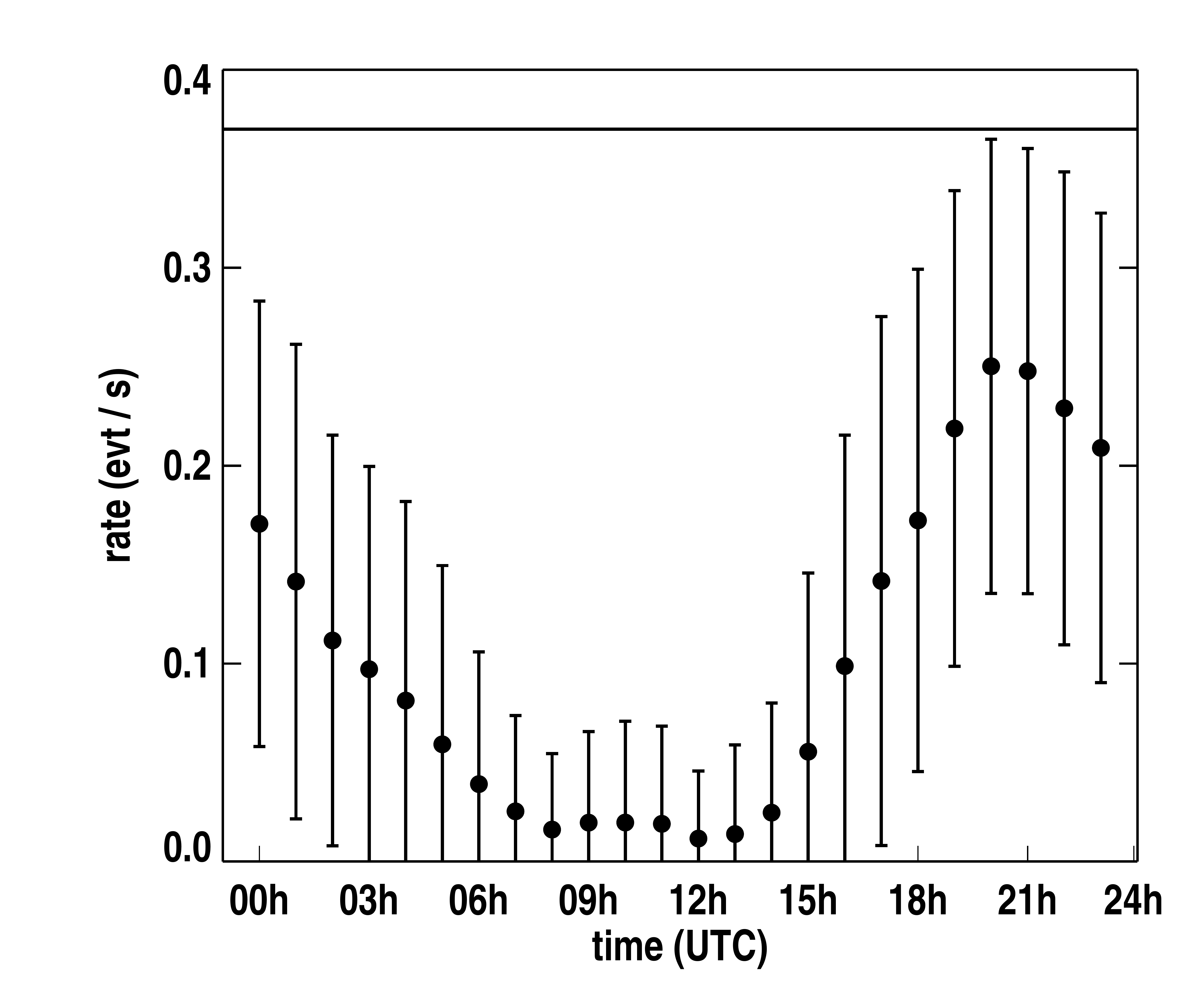}
\end{center}
\caption{\sl{Averaged weighted event rate (events per second) for A1 (where the weight is the number of events during each day of the sample) as a function of UTC time. The error bars indicate the $1\sigma$ value at each data point. The horizontal line indicates the saturation level at 0.37~event/s.}}
\label{fig:trig}
\end{figure}

Another way to present the trigger dependence on the noise conditions and to investigate its long-term behavior is to plot its average over a few months as a function of UTC hour. For station A1, this distribution is presented in Figure~\ref{fig:trig}, corresponding to $384\,000$ events between January 2008 and May 2008, where the trigger threshold was between 90 and 150~$\mu$V~m$^{-1}$ depending on the arrival direction of the event. As was already observed for the particular days displayed in Figure~\ref{fig:trig1}, the event rate is not uniform during the day. It starts to increase at 15~h~UTC, it reaches a maximum at 21~h~UTC and then slowly decreases until 9~h~UTC.

\begin{figure}[!ht]
\includegraphics[width=8cm,height=7.3cm]{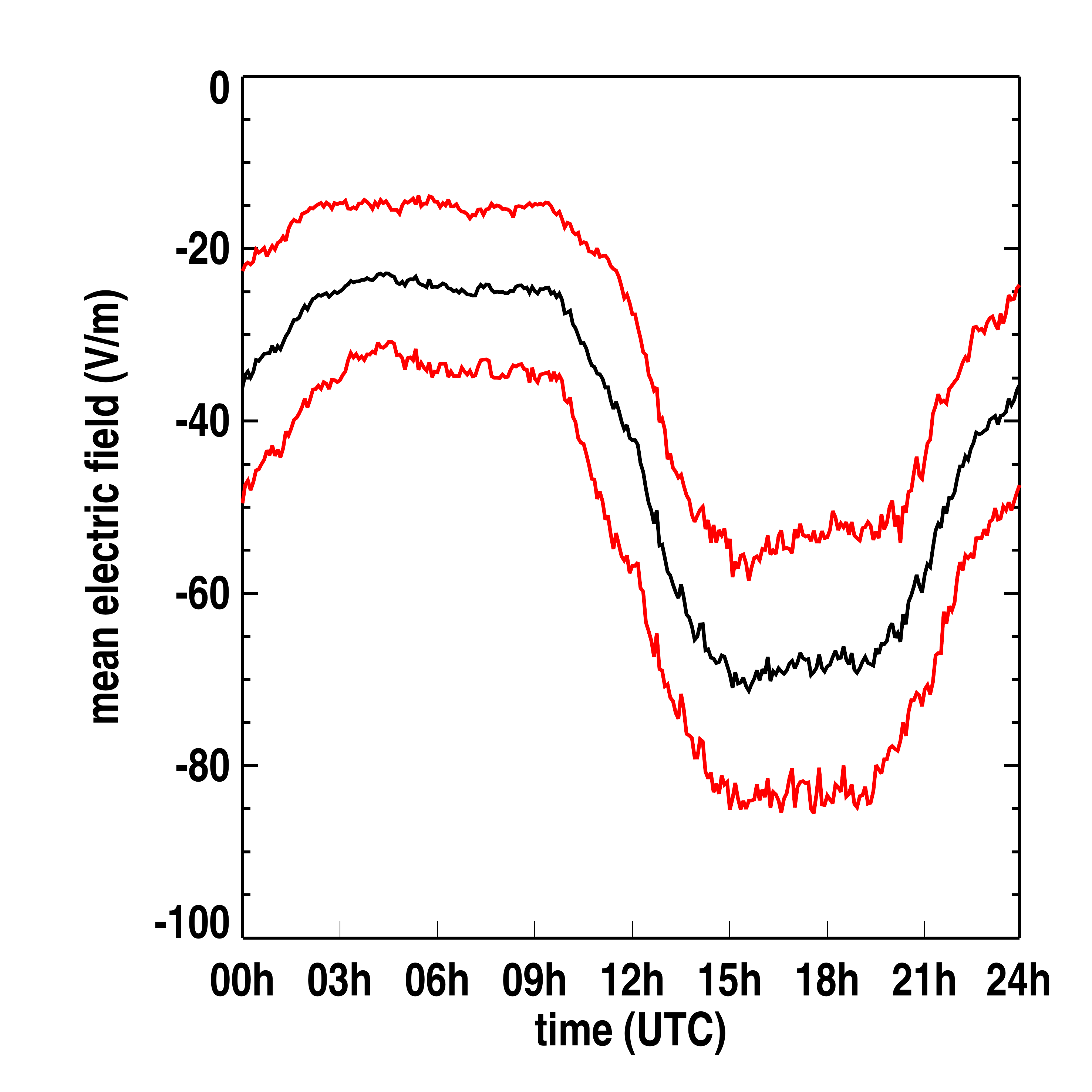}
\includegraphics[width=9cm,height=7.5cm]{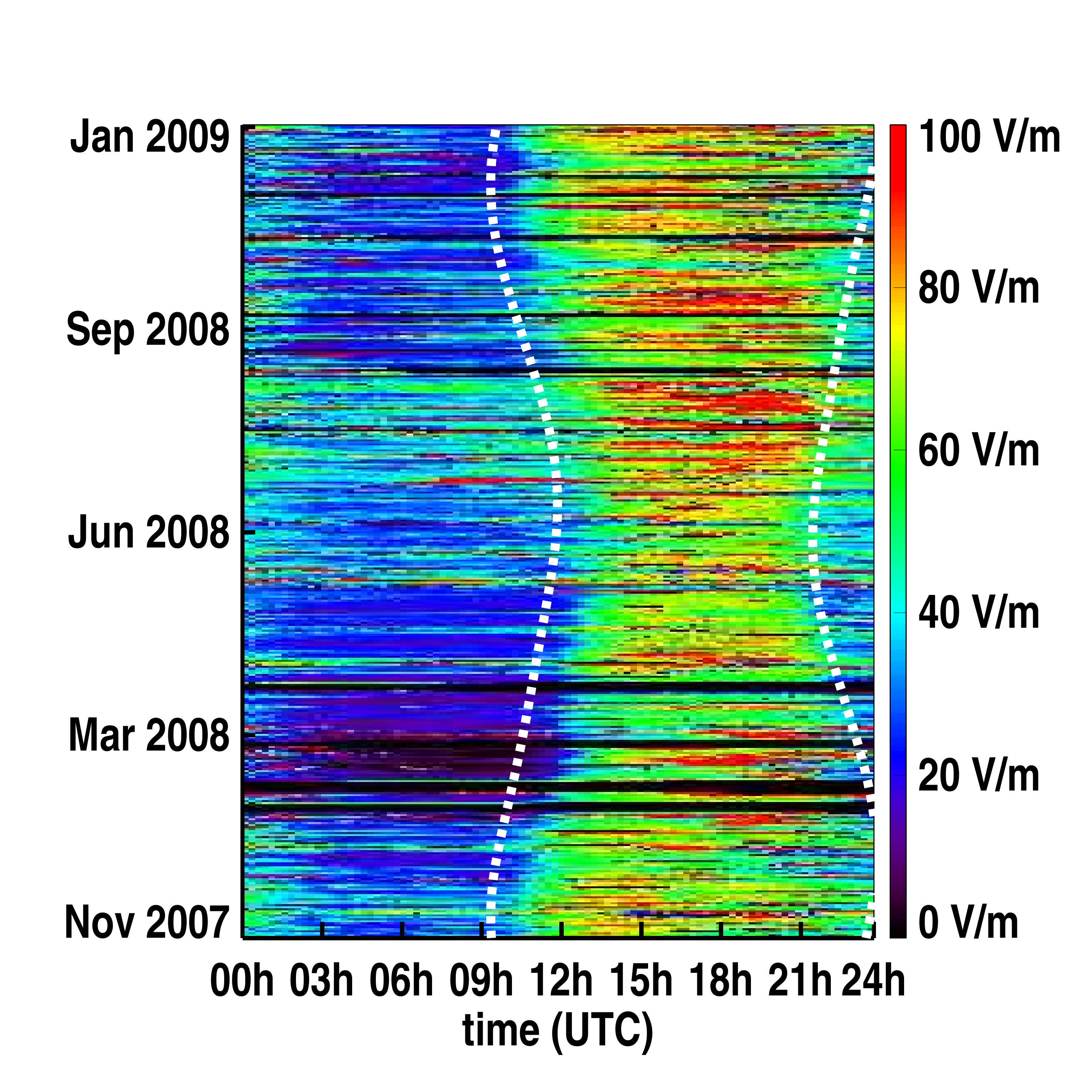}
\caption{\sl{Left: the daily variation of the static electric-field strength (averaged over one year: pathological values exceeding the average by more than 2~$\sigma$, like the ones recorded during thunderstorms, are not taken into account). The $\pm 1~\sigma$ range is represented. Right: the absolute value of the local electric field over 15 months (between November 2007 and January 2009). The yellow and green hourglass-shaped zone is due to the variation of the duration of the daily solar exposure. Black parts of this plot indicate missing data during very large perturbations of the electric field caused by close lightning hits. Hours of sunrise and sunset are superimposed as dotted-lines.}}
\label{fig:field}
\end{figure}

A possible origin for this daily variation could be the periodicity of the Earth's electric field strength. The local electric field was recorded by an electric-field meter located at the BLS (see Figure~\ref{fig:augerLayout}) which is about 18~km west of the site of this radio-detection array. This electric-field meter records every second the vertical static component of the electric field at ground level. The typical daily variation of the recorded values is presented in Figure~\ref{fig:field} (left) for values averaged over a period of one year.
The mean electric-field strength undergoes a periodic variation of about ($50 \pm 10$)~V~m$^{-1}$. This daily variation changes slightly with seasons and is correlated with the solar exposure, as shown in Figure~\ref{fig:field} (right).

\begin{figure}[!ht]
\begin{center}
\includegraphics[width=7.5cm]{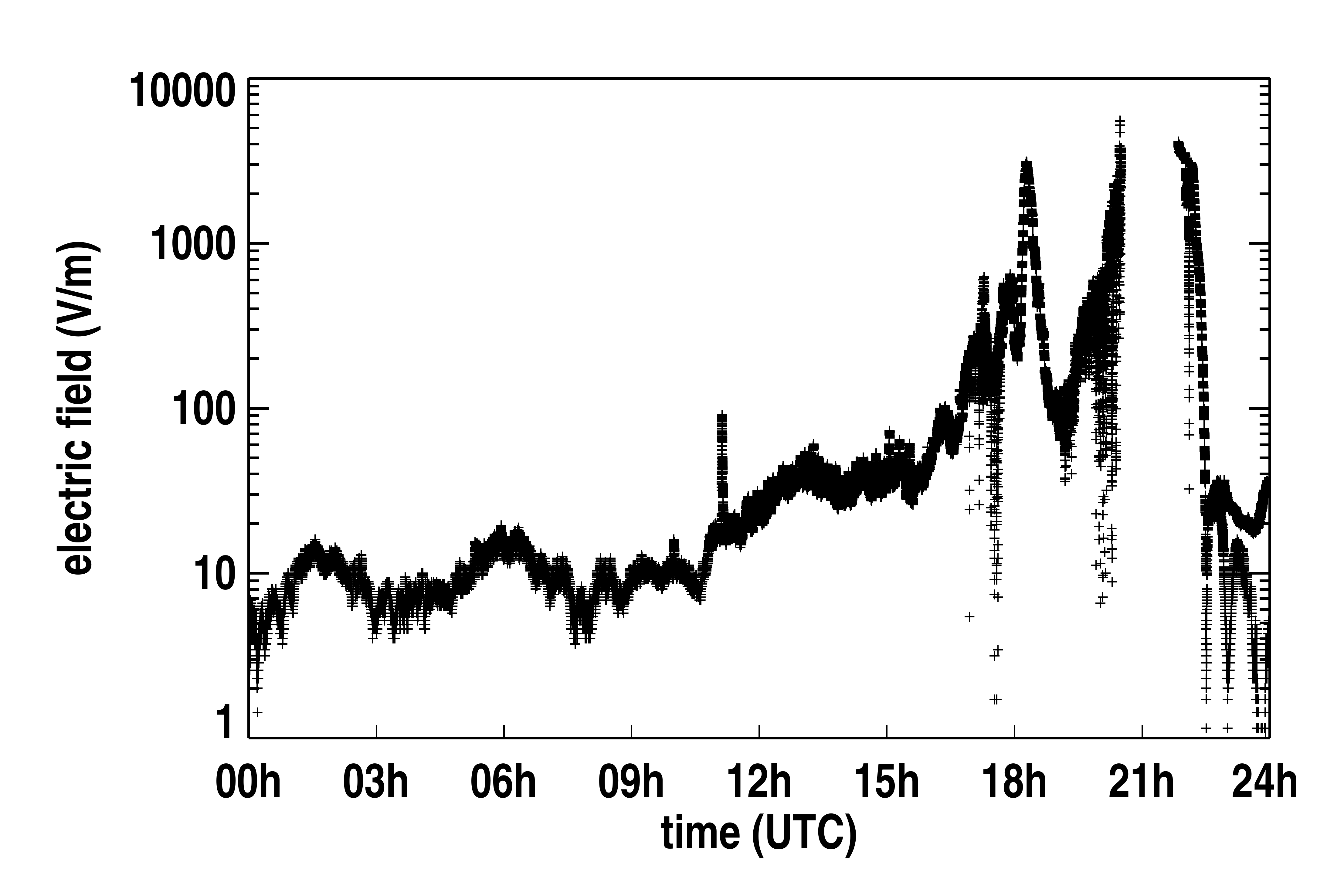}
\includegraphics[width=7.5cm]{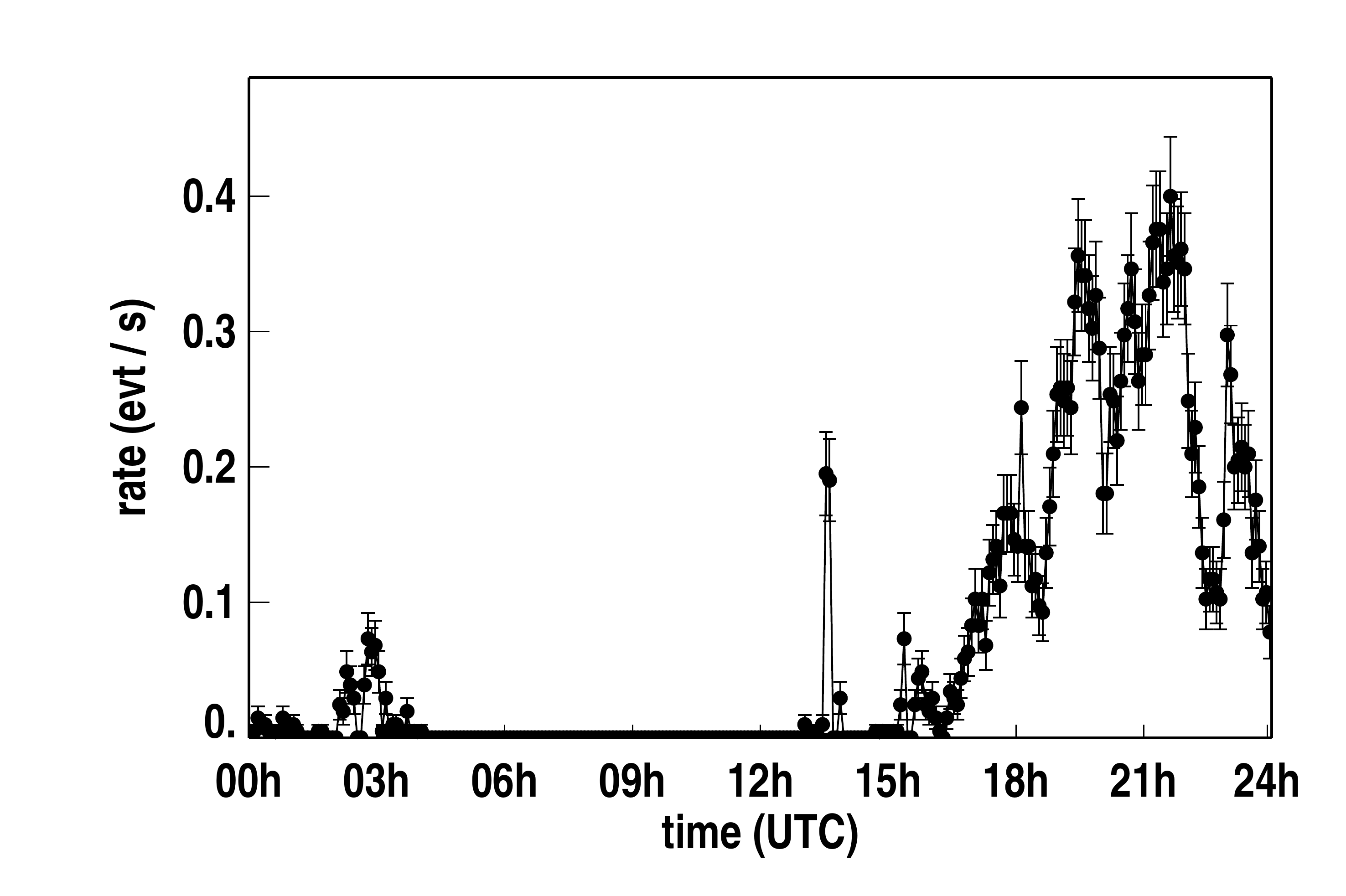}
\end{center}
\caption{\sl{Left: the recorded absolute value of the electric field on March 7, 2008. Right: the event rate for station A3 during the same day.}}
\label{fig:trfi}
\end{figure}

The measurement of the electric-field strength allows for the identification of, e.g., thunderstorms which cause very strong and wide-band transients possibly triggering a radio-detection station.
As an example, we display in Figure~\ref{fig:trfi} the event rate (right) for station A3 on March 7, 2008, which clearly follows the variation of the recorded strength of the static electric field (left).
From this observation, one can conclude that a strong deviation of the static component of the electric field also gives a strong counterpart in our trigger band (50-70~MHz).

\begin{figure}[!ht]
\begin{center}
\includegraphics[width=7cm,height=7cm]{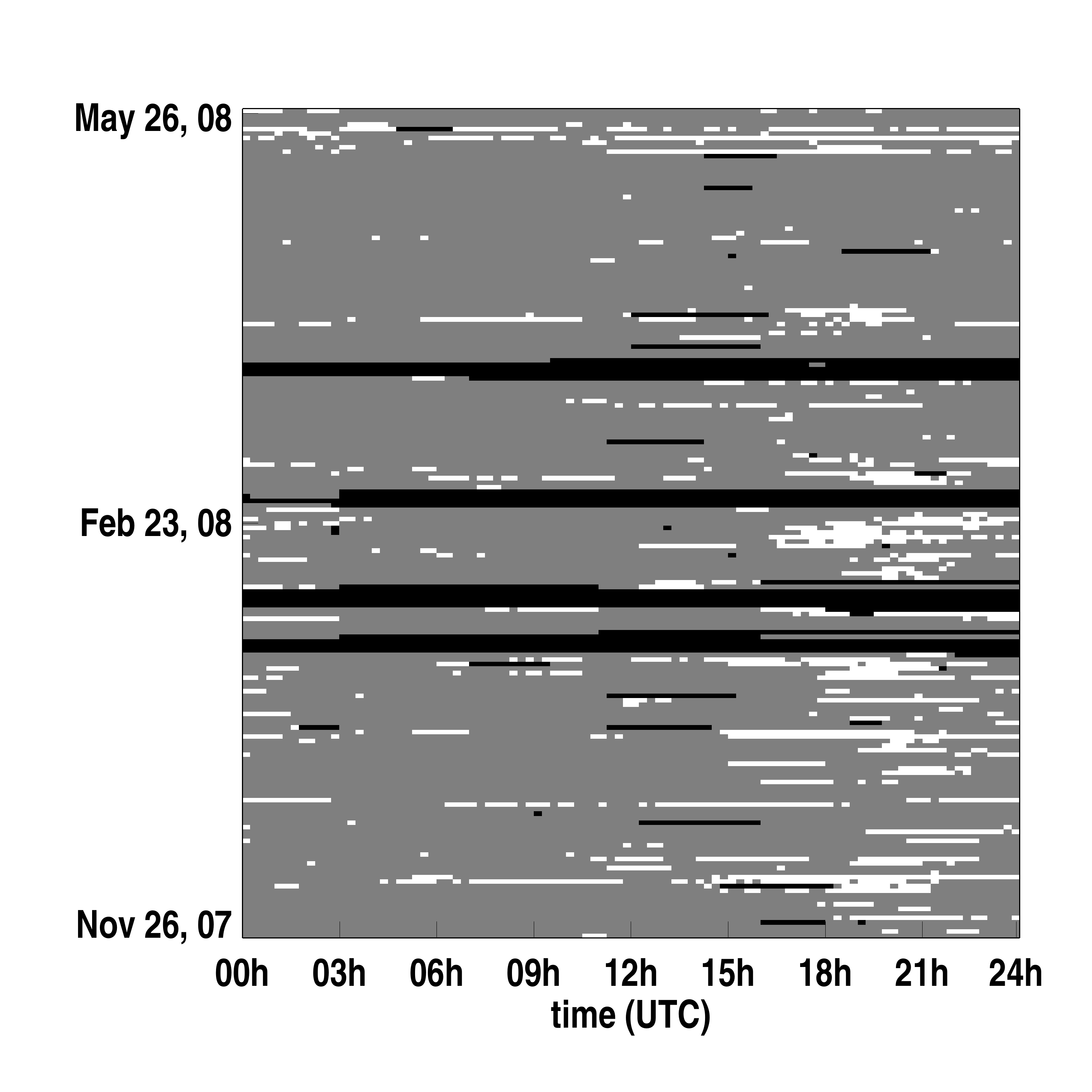}
\includegraphics[width=8cm,height=7cm]{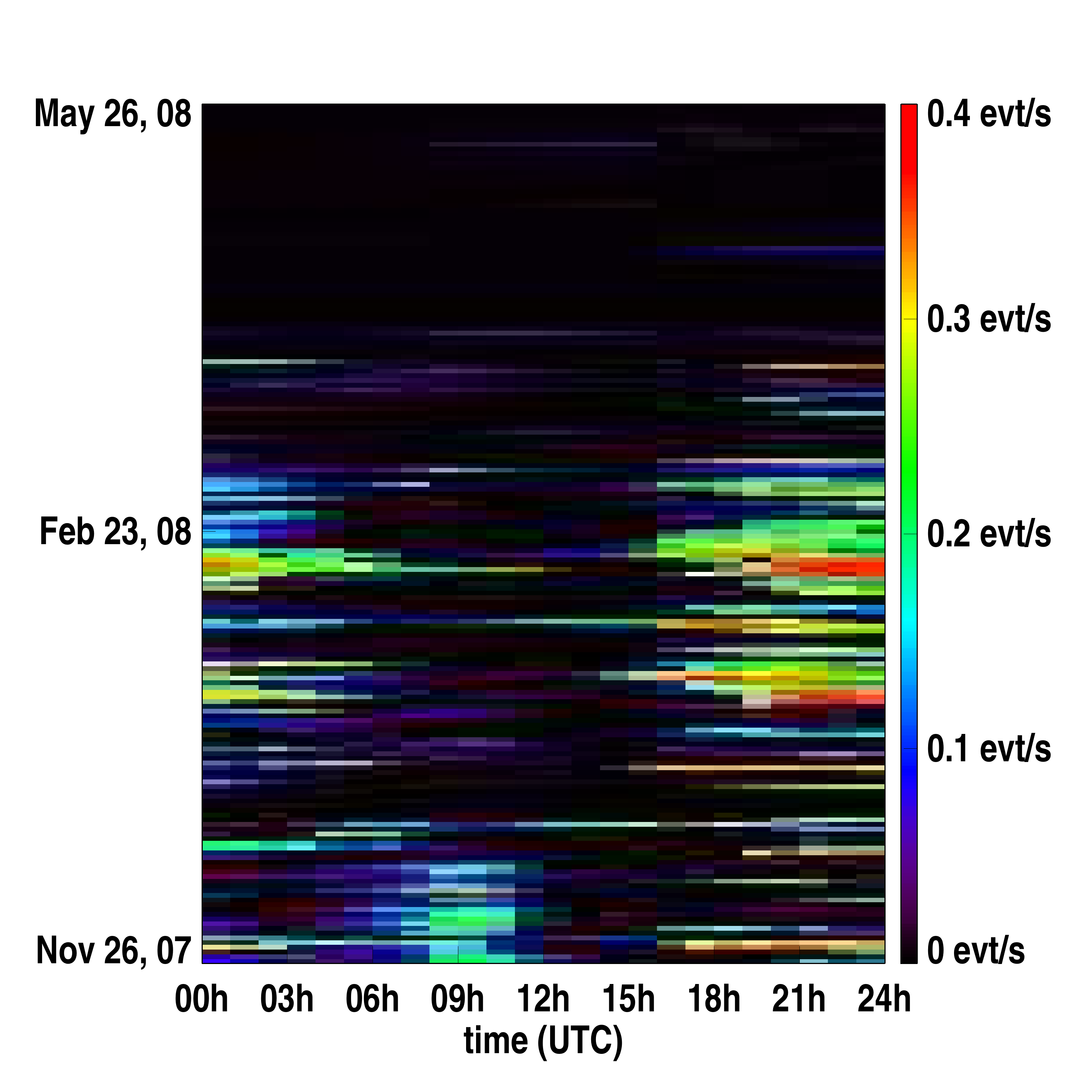}
\end{center}
\caption{\sl{Left: in white, recorded strong deviations (see text) of the value of the electric field as a function of time. Black parts of this plot indicate missing data during very large perturbations of the electric field caused by close lightning strikes. Right: recorded event rate for station A1.}}
\label{fig:trfi2}
\end{figure}

\subsection{Dependence on extreme electric-field conditions}

Now we assess periods of large static electric-field strength. The left panel of Figure~\ref{fig:trfi2} presents periods (in white) when the electric field deviates by more than 50~V~m$^{-1}$ from its mean absolute value. If we compare this plot with the one displayed in the right panel of this figure, we recognize that at the end of the day similar zones of high event rates and perturbations of electric field appear from December 2007 until the end of February 2008. Note that some electric-field data are missing (black zone in the left figure) during very large perturbations caused by severe thunderstorms.
 
\begin{figure}[!ht]
\begin{center}
\includegraphics[scale=0.3]{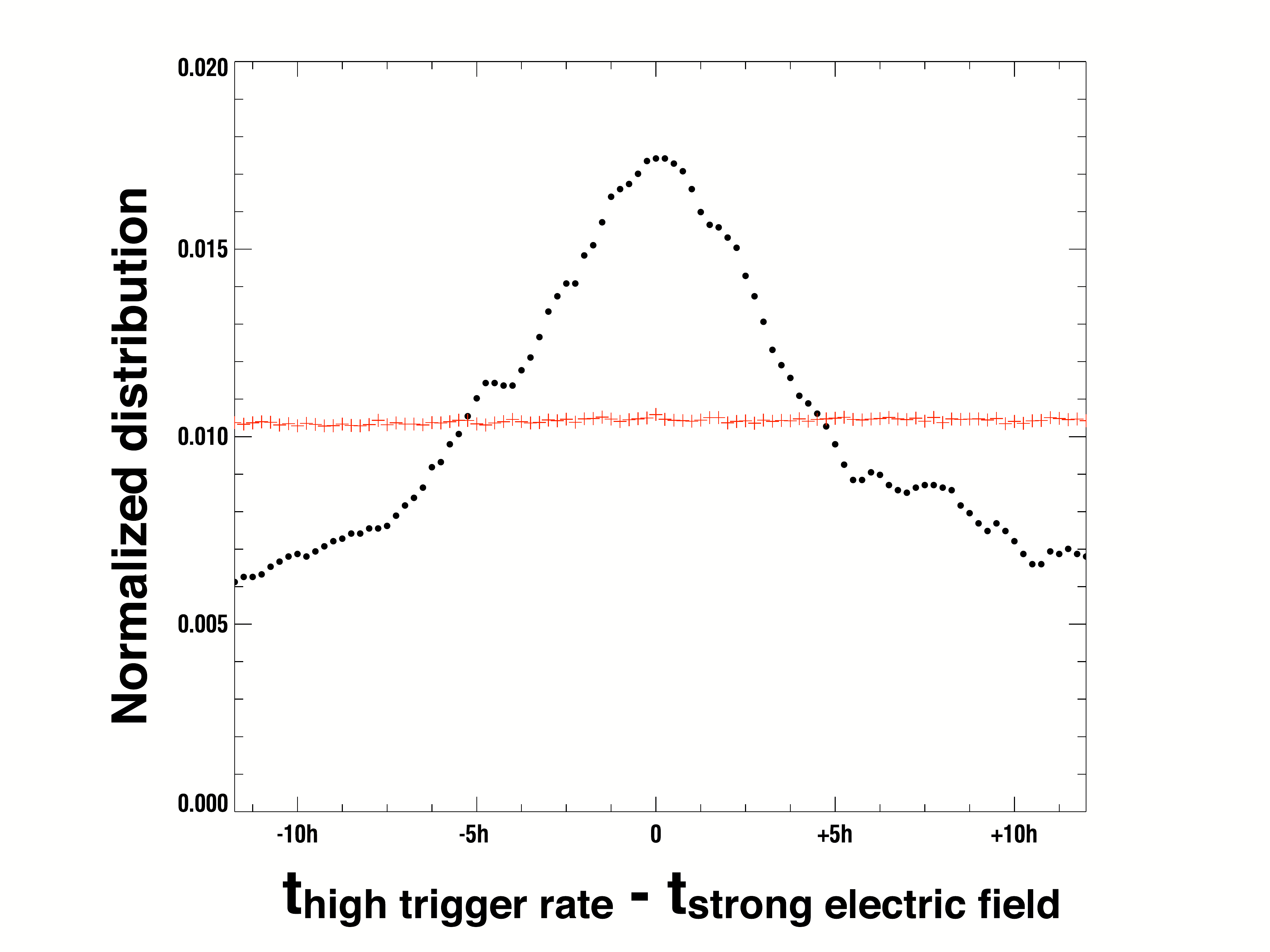}
\end{center}
\caption{\sl{Distribution of the time difference between saturation periods (more than 50 events within 15 minutes) for station A2 and periods with high values of the electric field (larger than 50~V~m$^{-1}$). The expected distribution in the case for which there is no correlation between the event rate and the strength of the electric field is shown as a horizontal line.}}
\label{fig:histoelec}
\end{figure}

The radio-detection setup is not able to detect transients induced by cosmic rays during periods of trigger saturation due to strong electric fields. Thus, all events detected with the radio-detection stations in coincidence with the SD have been recorded outside such periods.
For each time period where the event rate was larger than 50~events within 15~minutes (15\% of the saturation rate), we calculated all the time intervals $\Delta t$ between the considered time bin and the time bins (within $\pm 12$~hours) when the deviation of the electric field was larger than 50~V~m$^{-1}$. As an example, we display for station A2 the distribution of these time intervals in Figure~\ref{fig:histoelec}. The peak around $\Delta t\sim 0$ underlines the strong correlation between these two parameters. The same figure displays also the distribution one would have obtained in the case of no correlation between periods of a high event rate and periods with large values of the electric field.

\section{Coincident air-shower events}
\label{augerCoincidences}
As discussed in the previous section, the sensitivity of the radio-detection array varied according to the local weather conditions and the time of day. Therefore, the daily event rate varied strongly. Nevertheless, it is possible to identify easily actual cosmic-ray events in coincidence with the SD, due to the small value of the maximum event rate (0.37 event~s$^{-1}$). This guarantees that the number of random coincidences between radio and the SD is negligible.

\subsection{Independence and validation of coincident events}
\begin{figure}[!ht]
\includegraphics[width=8cm]{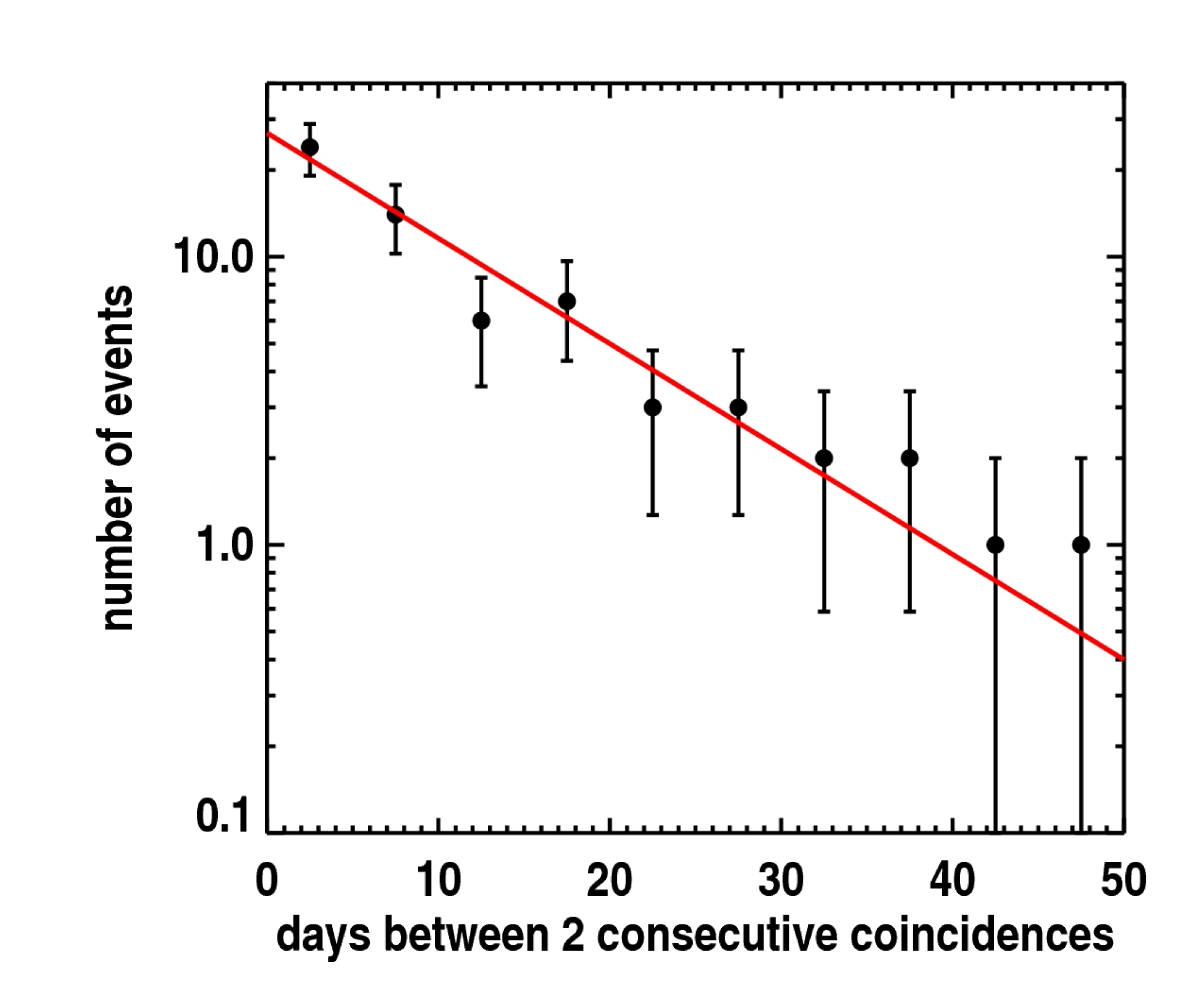}
\includegraphics[width=8cm]{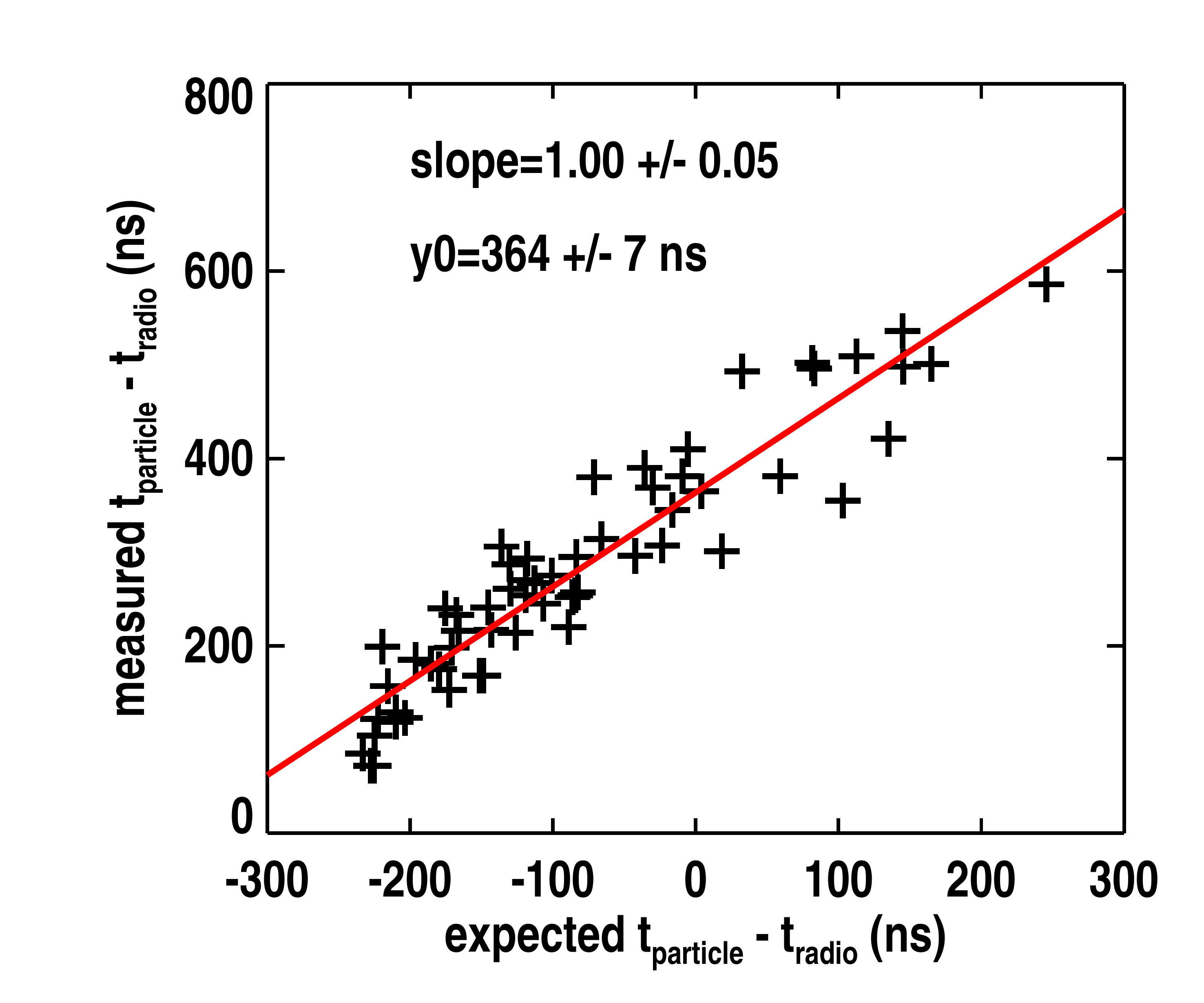}
\caption{{\sl Left: distribution of the time intervals between two consecutive events detected in coincidence by the SD and RAuger. Right: correlation between the measured time difference between the trigger time of Apolinario and RAuger, as a function of  the expected time difference given the shower geometry determined from the SD.}}
\label{fig:obsdt}
\end{figure}

We searched for coincidences between radio and the SD by an off-line comparison of the radio event times of each radio station with the arrival time of SD events computed at the location the SD station Apolinario. We used the SD T4 events with no fiducial cut (see~\cite{PieraTriggerPaper} for a definition of T4 events). If the difference in time between this SD event and the radio event is less than 1~$\mu$s, we verify that the radio event time is compatible with the geometry of the shower by comparing $t^{\mathrm{measured}}_{\mathrm{SD}}-t^{\mathrm{measured}}_{\mathrm{radio}}$ with $-(u\delta x+v\delta y)/c$, where $(u,v)=(\sin\theta\cos\phi,\sin\theta\sin\phi)$ are the shower axis direction cosines, and $\delta x=x_{\mathrm{SD}}-x_{\mathrm{radio}}$,  $\delta y=y_{\mathrm{SD}}-y_{\mathrm{radio}}$, with $(x_{\mathrm{SD}},y_{\mathrm{SD}})$ and $(x_{\mathrm{radio}},y_{\mathrm{radio}})$ the ground coordinates of an SD station and a radio station participating in the event, respectively. A conservative estimate of the instantaneous accidental coincidence rate in a time window of 20~$\mu$s gives a number of the order of $10^{-10}$~s$^{-1}$. For this calculation, we consider an average rate of 1.4 air showers per day registered by the SD within a range of 1~km from Apolinario, and we assume the worst situation where the radio event rate is always saturated. Therefore, the expected number of random radio events in coincidence with the SD in a conservative time window of 20~$\mu$s is of the order of $0.016$, integrated over the 2.6~years of running time of the experiment.
Up to May 2010, \nevts\ coincidences were recorded: 58, 6 and 1 with one, two or three radio stations, respectively. The distribution of the times between any two such radio-detected consecutive events is displayed in Figure~\ref{fig:obsdt} (left). The solid line in this figure describes an exponential fit to the data with a time constant of $11.8\pm 0.2$~days.
To investigate whether these radio events were not generated by other sources (e.g., the electronics or the photomultipliers of nearby SD stations), the difference in the triggering time from the radio stations participating in the event and that from the SD station Apolinario was calculated. In Figure~\ref{fig:obsdt} (right), we compare these time differences with the predicted values based on the shower geometry as determined from the recorded SD data.
The correlation appears to be very clear: the slope of the linear fit through the data points is equal to one.
 
\subsection{Energy and distance distributions for coincident events}\label{distributions}
\begin{figure}[!ht]
\includegraphics[scale=0.28]{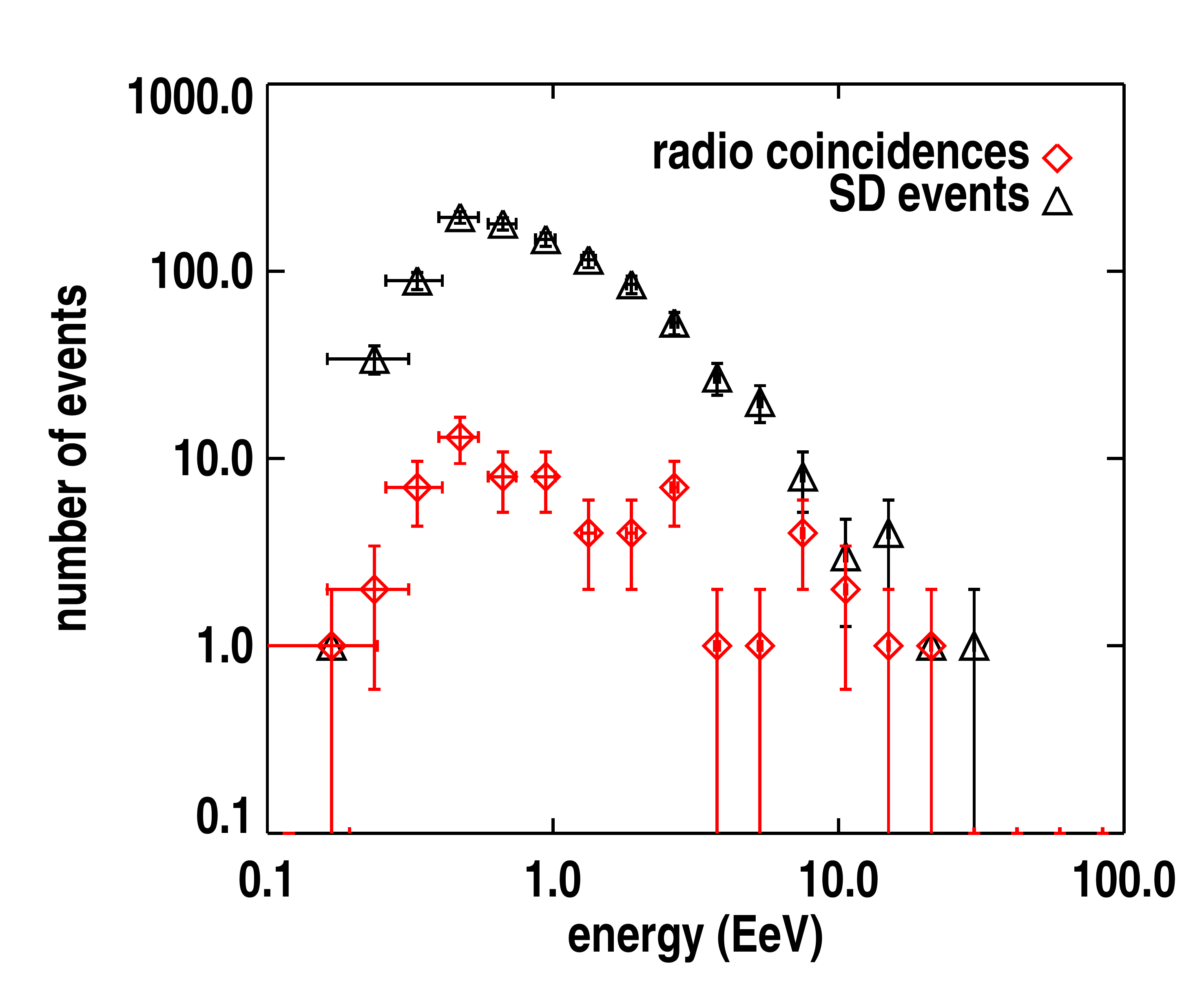}
\includegraphics[scale=0.28]{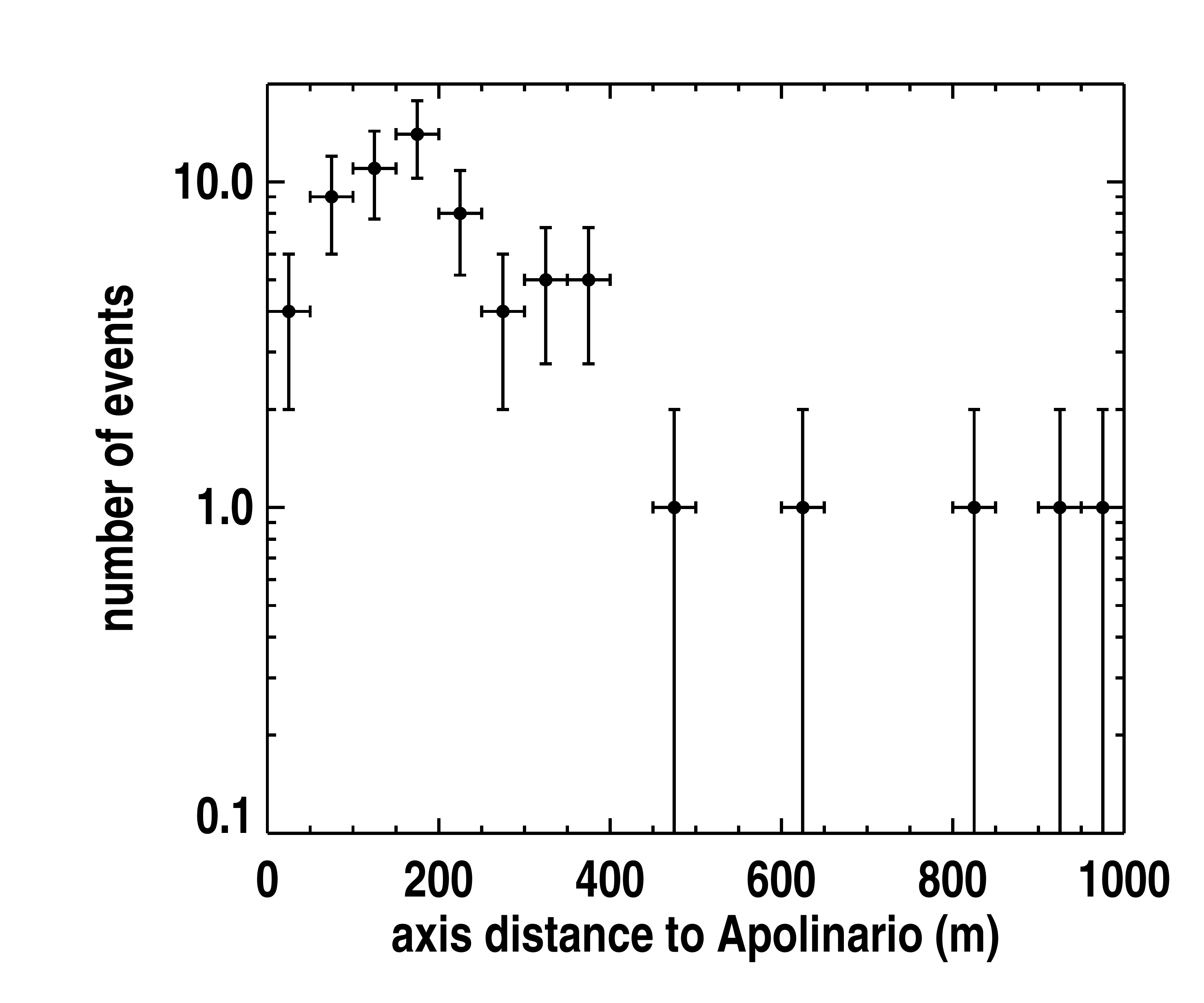}
\caption{\sl{Left: the energy distribution of the coincident events compared to the complete set of SD events in the same time period, having a shower axis distance to Apolinario smaller than 1~km. Right: the distribution of the distance between the shower axis and Apolinario for the coincident events.
}}
\label{fig:distributions}
\end{figure}

Figure~\ref{fig:distributions} (left) shows the energy distribution of the \nevts\ registered coincident events, compared to the set of all SD T4 events that 1) could have been detected by any of the 3 radio-stations, and 2) had a shower axis distance to Apolinario smaller than 1~km.
The total number of SD T4 events is 962, with no cuts on the reconstructed arrival directions and energies.
In this SD reconstruction analysis, the data registered by Apolinario were discarded, because this station was not part of the standard SD grid.
The shapes of the two distributions shown in the figure are compatible in the interval 0.8-2~EeV: the slopes of the energy distributions are $-2.1\pm  0.3$ and $-1.9\pm 0.9$ for the SD event and radio-coincidences, respectively. Given the small number of coincident events, we cannot give a conclusion with a good confidence level on the radio energy threshold for this prototype.

\begin{figure}[!ht]
\begin{center}
\includegraphics[width=12cm]{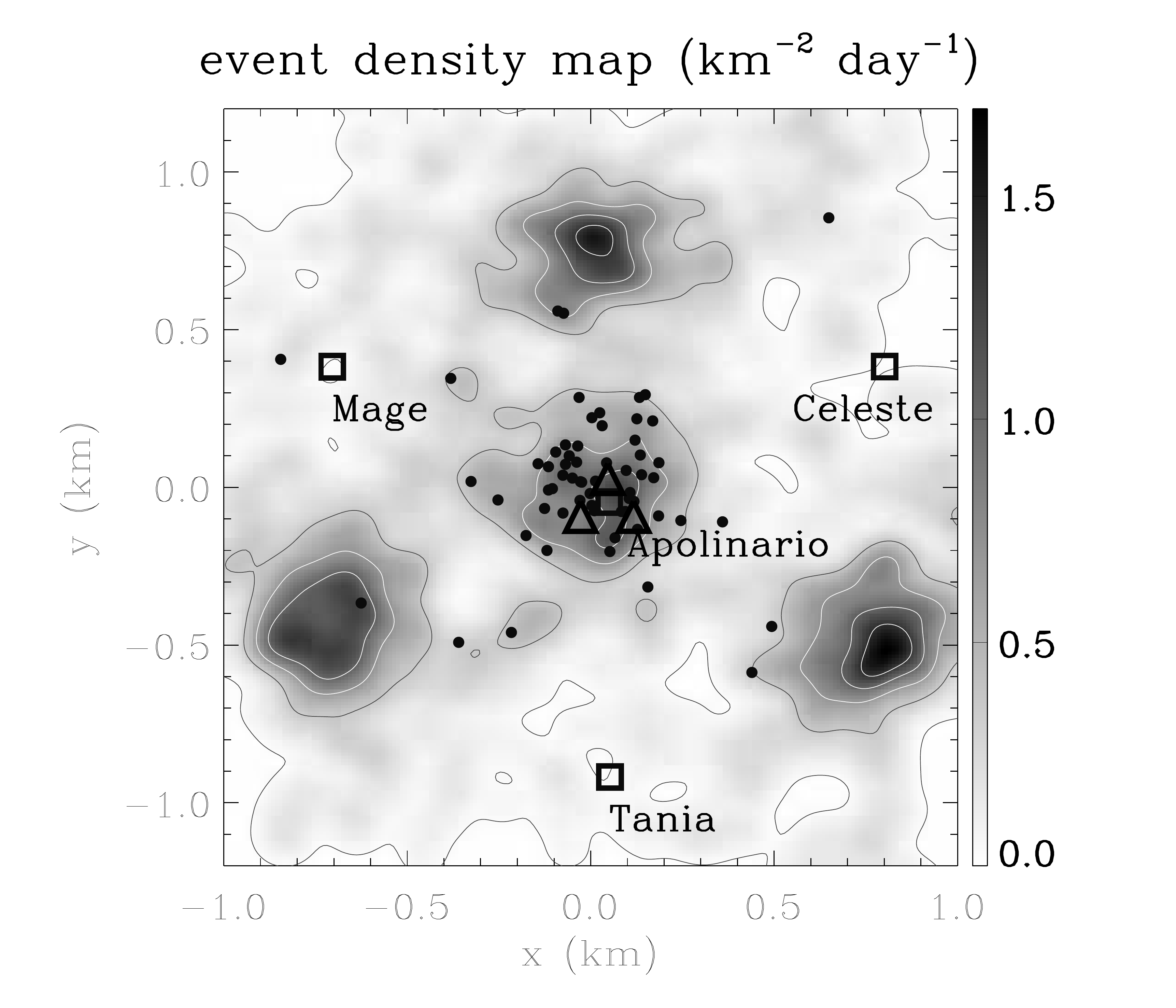}
\end{center}
\caption{\sl{Density map (events km$^{-2}$ day$^{-1}$) of the position of the reconstructed shower cores as determined from the SD during the whole period of the radio observations. The enhanced densities correspond to the centers of the elementary SD triangles. The black dots in this plot are the coincident events; locations of the SD stations are marked as the black squares. Radio stations A1, A2 and A3 are the black triangles close to Apolinario.}}
\label{fig:densityMap}
\end{figure}

The right panel of Figure~\ref{fig:distributions} displays the distribution of the shower axis distance from Apolinario for the coincident events. 60 events of the \nevts\ coincident events have an axis distance smaller than 400~m, with 2 events recorded at more than 900~m. Note that these 2 events are remarkable because they are the most inclined events in coincidence, with zenith angles of $75.5^\circ$ and $78.5^\circ$. We compute that the detection efficiency of these inclined events with axis distances above 900~m is 50\%, which is a much higher value than for the whole set of coincident events. The relative detection efficiency is discussed in section~\ref{deteff}. Figure~\ref{fig:densityMap} displays the density map of the core positions, reconstructed from the set of 962~SD events. Note that each core position is smoothed with a 2D Gaussian spread function of 50~m width. In addition, we show in this plot the positions of the coincident events, which follow clearly the density map determined from the SD events alone. Again, we conclude that the coincident events have no bias with respect to the normal SD events.



\subsection{Relative detection efficiency for coincident events}\label{deteff}
In this section we discuss the relative detection efficiency of RAuger with respect to the SD. To do so we use an event selection requiring that the detector station with the highest signal be surrounded by operational stations, i.e., T5 events~\cite{PieraTriggerPaper}.

In the period over which the radio-detection array was deployed,
the effective running time fractions for A1, A2, and A3 were 59\%, 42\%, and 43\%, respectively, taking into account the time periods when the event rate is less than 67\% of the saturation.
Here we note that these fractions are overestimated, because  not all break-down periods of these stations have been recorded.

\begin{figure}[!ht]
\begin{center}
\includegraphics[width=14cm]{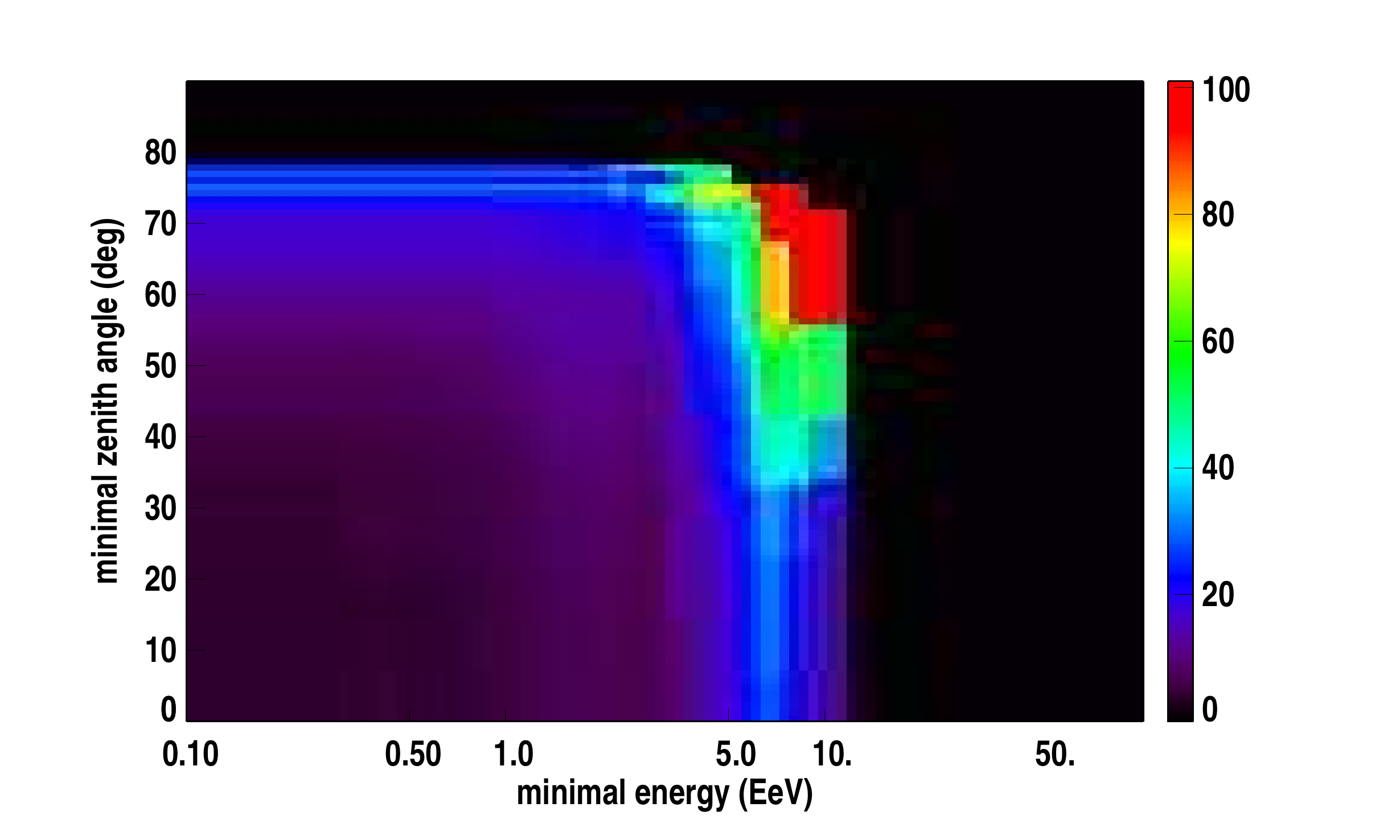}
\caption[]{{\sl Relative detection efficiency for station A1 as a function of minimal zenith angle and minimal energy. The color scale is in percent. Note that the maximum zenith angle of the coincident events is $78.5^\circ$ and the maximum energy is 12.3~EeV; this explains the black regions above these values.}}\label{fig:deteff}
\end{center}
\end{figure}

A1, A2 and A3 detected a total of 35, 8 and 4 showers in coincidence with the SD, respectively and could have observed 908, 681 and 714 showers, respectively (estimate based on their on-time). Therefore, the raw relative efficiency with respect to the SD events is 3.9\%, 1.2\% and 0.5\%, respectively. Nevertheless, it is interesting to check the influence of energy and zenith angle on the relative efficiency. Figure~\ref{fig:deteff} shows this efficiency for station A1 with respect to the SD as a function of a cut on minimal zenith angle and minimal energy. 
Considering the relative detection efficiency for station A1 as a function of minimal zenith angle only, shows a clear increase of the efficiency with increasing zenith angle. For instance, the efficiency reaches a maximum of 30.8\% for $\theta\geqslant 75^\circ$. In the same way, considering the relative detection efficiency as a function of minimal energy only, shows a clear increase of the efficiency with the minimal energy, and its maximal value is 28.6\%. The effect is weaker when considering the minimal axis distance only, we find in this case that the detection efficiency is 3.9\%. 
RAuger is therefore relatively more sensitive to inclined and high-energy showers.

\subsection{Arrival directions of coincident events}\label{geomagnetic}
\begin{figure}[!ht]
\centering
\includegraphics[scale=0.5]{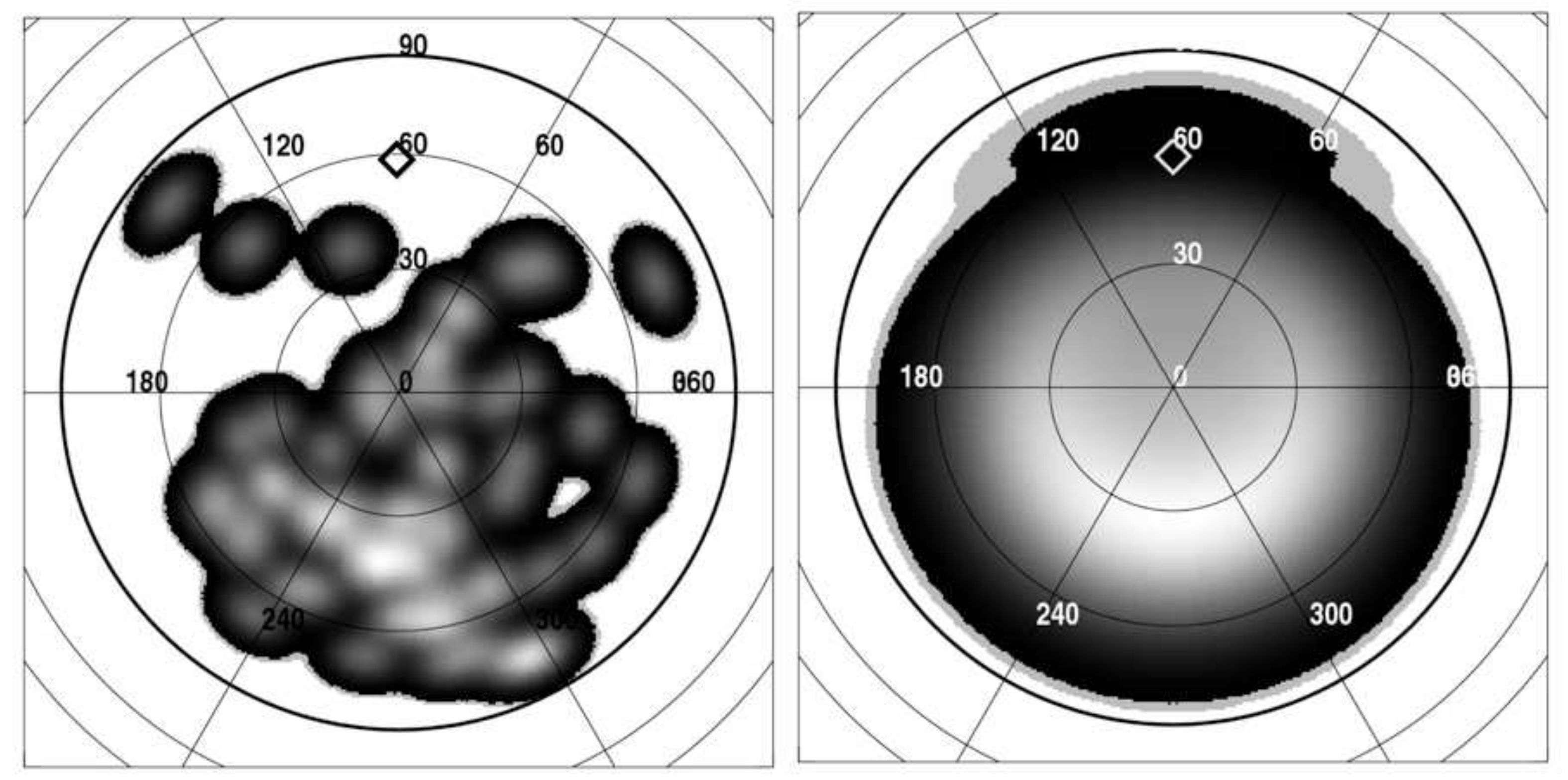}
\caption{\sl{Left: the sky-map distribution (normalized to the maximum density) in local coordinates of the \nevts\ coincident events. Right: the predicted sky-map distribution based on the geomagnetic model (same color code). Both distributions have been smoothed by a $10^\circ$ Gaussian spread function. The diamond at $(\theta=58^{\circ},\phi=90^{\circ})$ indicates the direction of the geomagnetic field in Malarg\"ue.}}
\label{fig:skymap}
\end{figure}

\begin{figure}[!ht]
\begin{center}
\includegraphics[width=8cm]{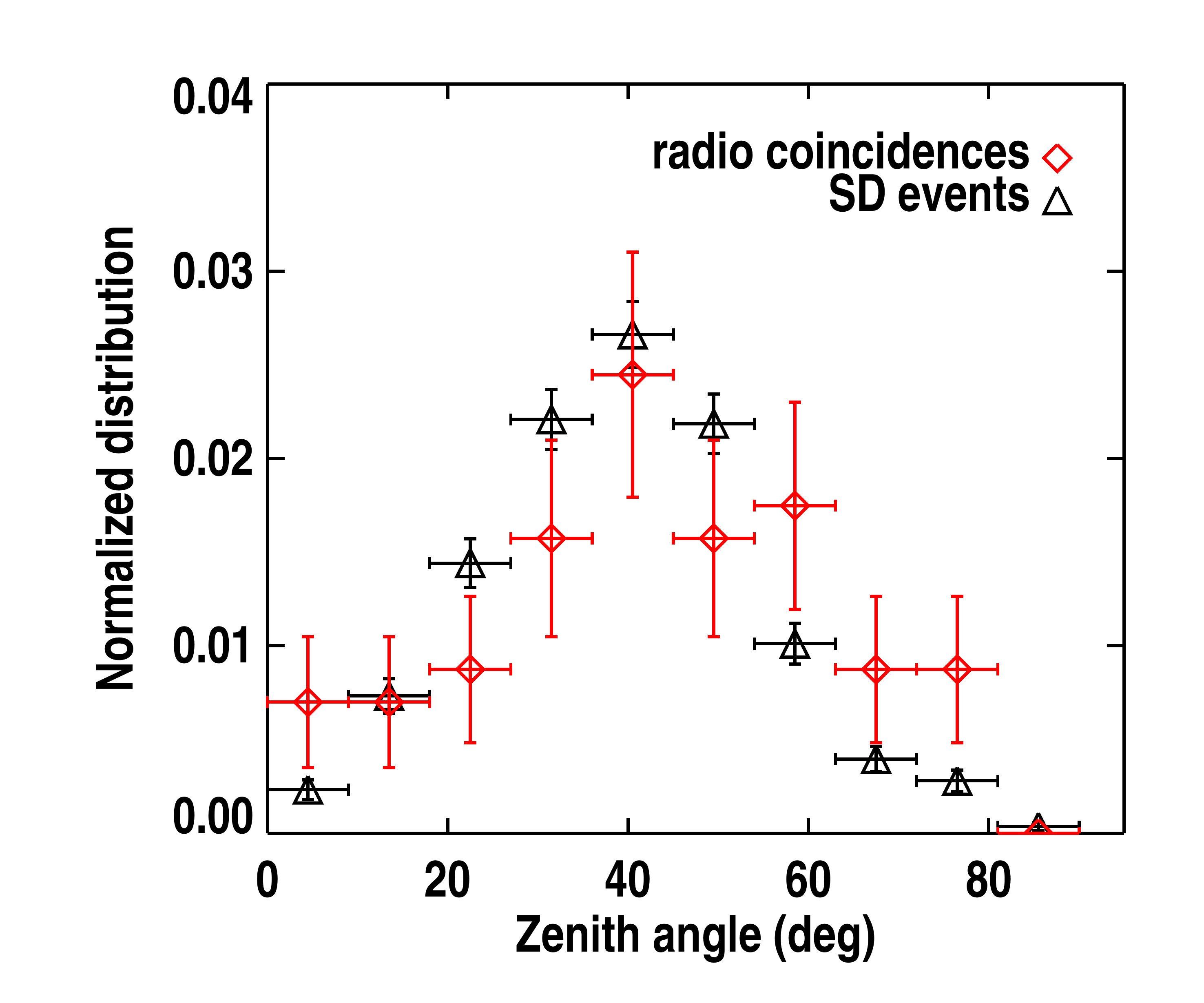}
\includegraphics[width=8cm]{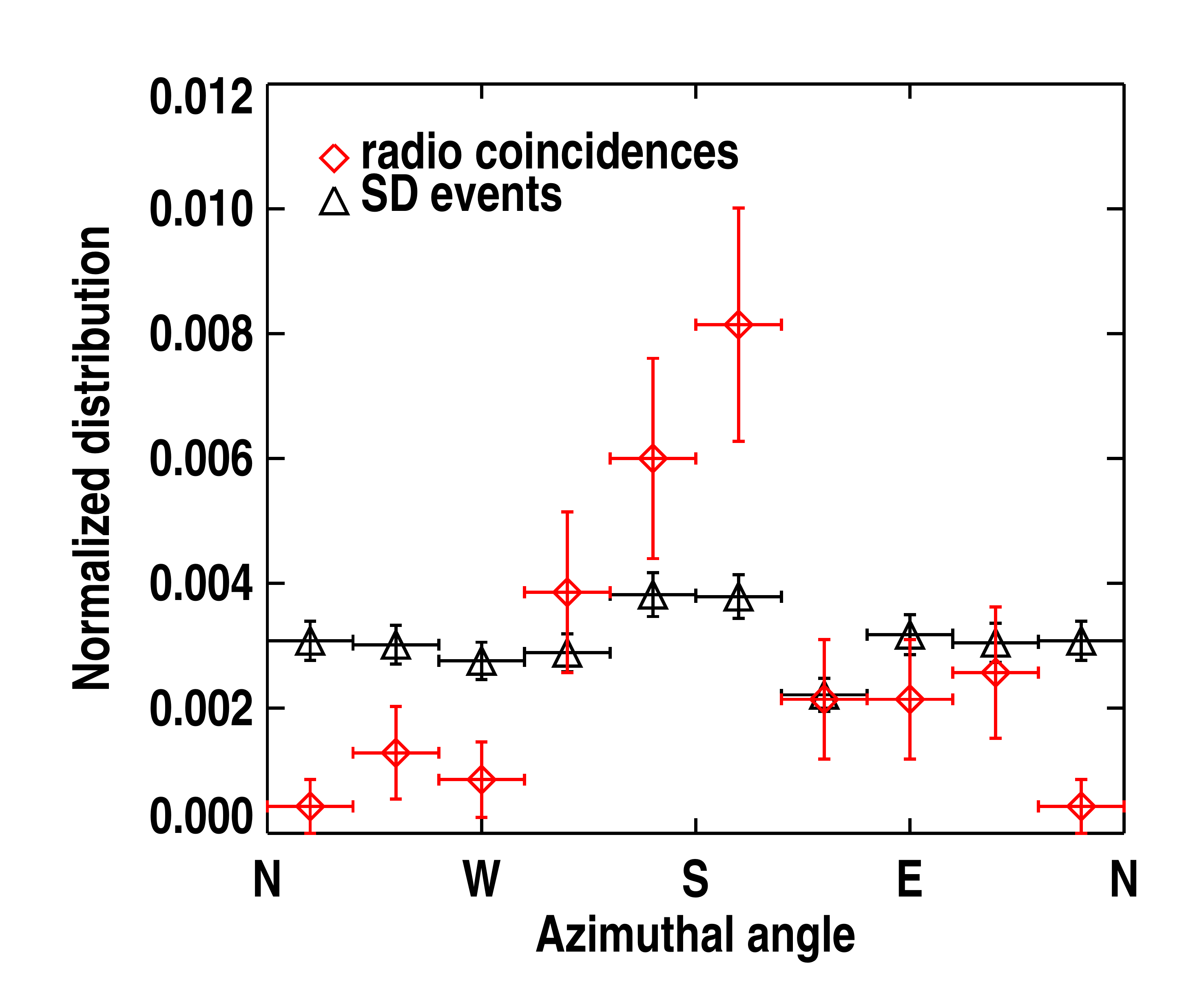}
\caption{{\sl Comparison of the normalized angular distributions as a function of the zenith angle $\theta$ (left) and azimuthal angle $\phi$ (right) for the coincident events and SD-only events.}}
\label{fig:angularDistributions}
\end{center}
\end{figure}

The distribution of the arrival directions of cosmic rays, recorded by a radio-detection setup, can be described fairly well using a simple geomagnetic emission model, as shown by the CODALEMA experiment~\cite{coda09}.
In this model, the induced electric-field strength {\boldmath$\mathcal{{E}}$} has the form  {\boldmath$\mathcal{E}\propto \mathbf{n}\times\mathbf{B}$}, where $\mathbf{n}$ indicates the direction of the shower axis and $\mathbf{B}$ describes the geomagnetic field; see for instance Refs. \cite{kahnLerche,scholtenwerner}.
The detection probability is proportional to the induced electric-field strength, which can be exploited to predict an event density distribution, as done in~\cite{coda09}. Such a map has been computed for the Malarg\"ue site, assuming an isotropic arrival distribution of cosmic rays and using the local orientation and strength of the geomagnetic field.
This distribution was then multiplied by the EW projection of the $\mathbf{n}\times\mathbf{B}$ vector, because we were triggering on the EW polarization only.
The corresponding density map, smoothed with a $10^\circ$ Gaussian spread function, is presented in the right panel of Figure~\ref{fig:skymap}; in the left panel we display the observed sky-map distribution, using the same spread function.
The asymmetry in the arrival directions of detected events is shown in Figure \ref{fig:angularDistributions} as normalized angular distributions. The angular distributions as a function of $\theta$ (left panel) and $\phi$ (right panel) for the coincident events are shown together with those from the SD-only events, recorded during the same time period and with a distance from the shower axis to Apolinario which was less than 1~km. A large excess appears for the coincident-event distribution for directions coming from the south (82\% of the total number of events), while the two angular distributions as a function of the zenith angle $\theta$ are similar. We note that the detection of inclined showers by RAuger is relatively more efficient as previously stated in section~\ref{deteff}.

\begin{figure}[!ht]
\begin{center}
\includegraphics[width=8cm]{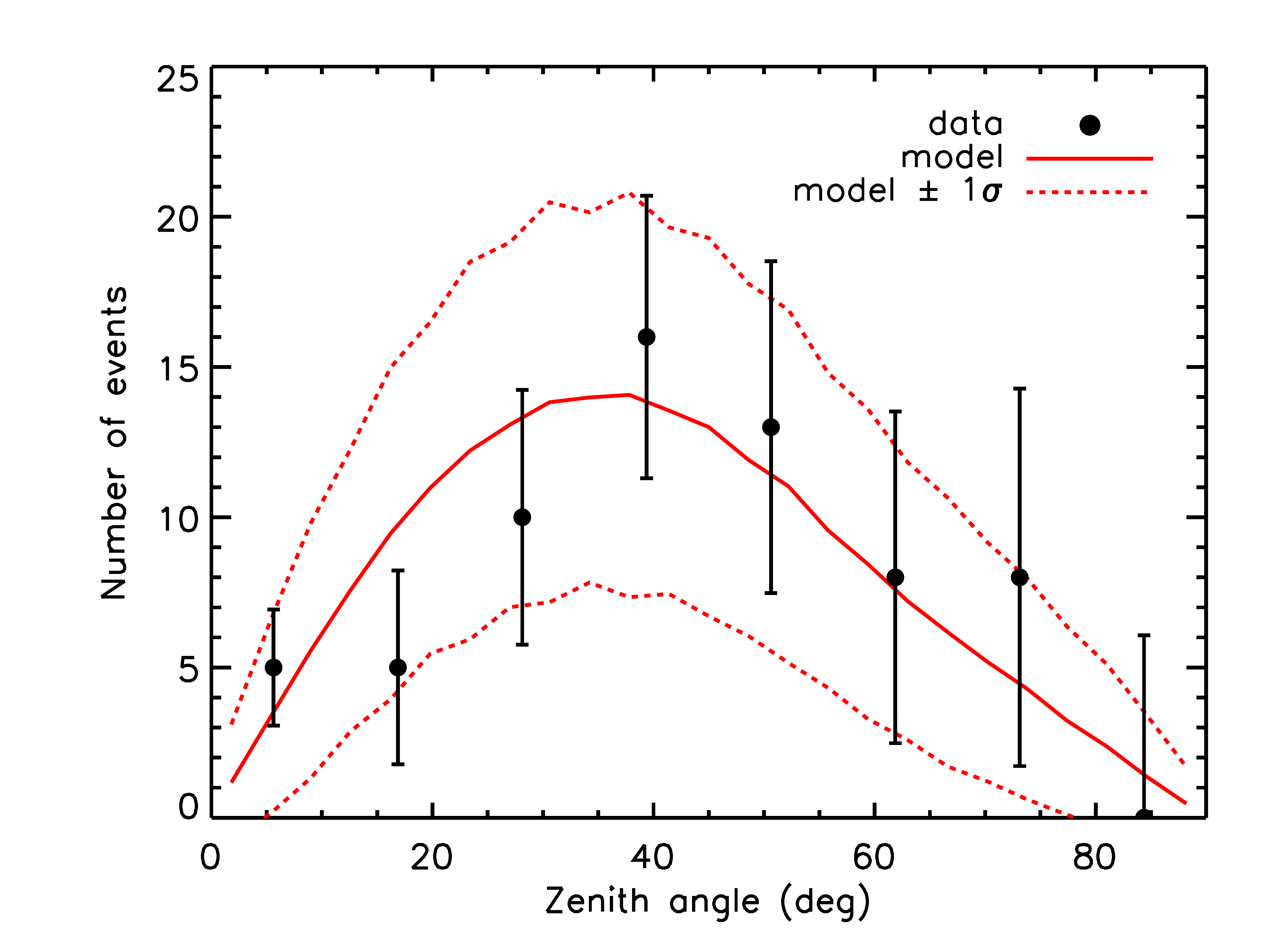}
\includegraphics[width=8cm]{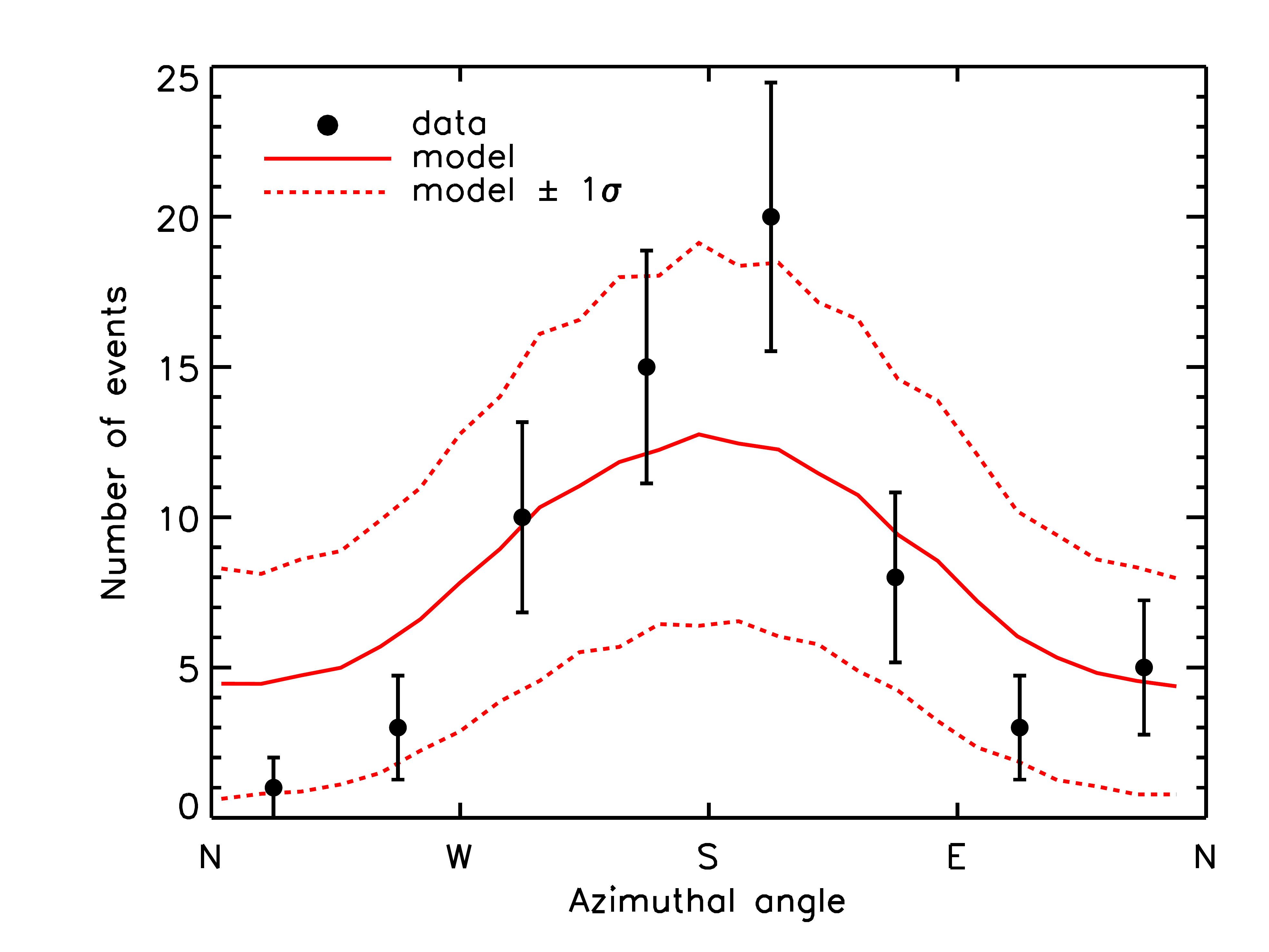}
\caption[]{{\sl Comparison of the expected zenith (left) and azimuthal (right) angular distributions with the observed ones. The solid lines correspond to the ensemble-averaged simulated angular distributions obtained after $10\,000$ realizations of \nevts\ events with the associated $\pm 1~\sigma$ error bands (dotted lines) following the predicted density map displayed in the right panel of Figure~\ref{fig:skymap}.
The points with error bars correspond to the measured data.}}
\label{fig:geomagneticTest}
\end{center}
\end{figure}

A further test of the assumed geomagnetic model can be performed by computing an ensemble average of the angular distributions
for a large number ($10\,000$ in this case) of realizations of $N$ simulated events following the expected density map (see the right panel of Figure~\ref{fig:skymap}), where $N=\nevts$ is the actual number of recorded coincident events. The observed zenith and azimuthal distributions are shown in Figure~\ref{fig:geomagneticTest} together with the angular distributions of the simulated events. Note that the simulated angular distributions were not fitted to the observed angular distributions. The agreement is satisfactory for both distributions and we can reproduce the excess of events coming from the south.
This confirms the dominant role of the geomagnetic field in the emission process.

\subsection{Correlation between shower energy and electric field}\label{correlation}
\begin{figure}[!ht]
\begin{center}
\includegraphics[width=17cm]{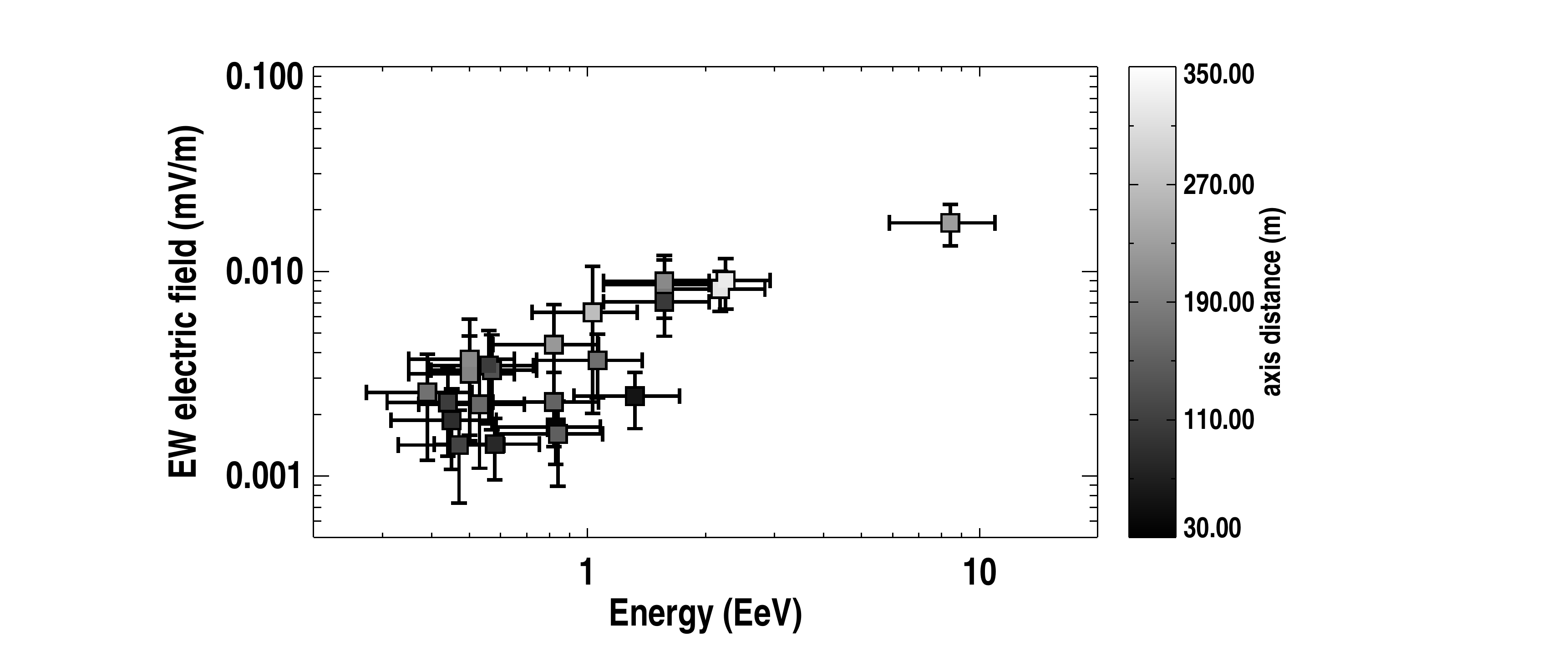}
\end{center}
\caption{{\sl Correlation between the primary energy of the showers and the electric-field
value calculated on the shower axis assuming an exponential decrease of the electric-field strength.
Vertical error bars are computed event by event and horizontal error bars ($\pm 30\%$) reflect the uncertainty of the reconstructed primary energy. The grey scale indicates distance to shower axis.}}
\label{fig:energy}
\end{figure}

The following considerations use the measurements in the EW polarization only because the signal is stronger than in the NS polarization. 
Historically~\cite{allan}, an exponential dependence of the amplitude of the electric field with axis distance has been used (written here for the EW polarization): ${\mathcal{E}^{\mathrm{EW}}(d)=\mathcal{E}^{\mathrm{EW}}_0\, e^{-d/d_0}}$ where $\mathcal{E}^{\mathrm{EW}}_0\ $ is the electric field amplitude at the shower axis and $d$ is the distance between an antenna and this axis. Most of our events in coincidence with the SD were detected by only one station (A1). Therefore, the direct estimation of both $\mathcal{E}^{\mathrm{EW}}_{0}$ and $d_{0}$ for each event is not possible. Nevertheless, it is still possible to obtain an estimation of the electric field value $\mathcal{E}^{\mathrm{EW}}_{0}$, assuming the parameter $d_0$ to be 150~m as suggested by the CODALEMA~\cite{codad0} and LOPES \cite{lopesd0} experiments. We also have to rescale $\mathcal{E}^{\mathrm{EW}}_{0}$ to a normalized value depending on the incoming arrival direction and the geomagnetic field $\mathbf{n}\times\mathbf{B}$. Finally, our estimator is given by ${\mathcal{E}^{\mathrm{EW}}_{0}=\mathcal{E}^{\mathrm{EW}}(d)\exp(d/d_0)/|(\mathbf{n}\times\mathbf{B})_\text{EW}|}$.
The resulting correlation between $\mathcal{E}^{\mathrm{EW}}_{0}$ and the SD estimation of the shower energy ${E}_{\mathrm{SD}}$ is presented in Figure~\ref{fig:energy}.
The correlation between the electric field detected by A1, A2 or A3 (filtered in the 40-80~MHz band) and the shower energy is visible. The error bars on the electric-field strength are computed by Monte Carlo calculations, propagating for each event the uncertainties on the shower geometry (core position, azimuth and zenith angle) and assuming a 20\% systematic error due to the soil conditions uncertainty. A power-law dependence is found but no calibration curve can be extracted from these data given the assumption made on the parameter $d_{0}$. Note that the SD events used to obtain this correlation have a zenith angle below $60^\circ$ and are selected according to the fiducial cuts required in all Auger SD analyses (T5 events, see~\cite{PieraTriggerPaper}). The number of resulting events is 19, corresponding to 23 radio traces. The raw Pearson correlation coefficient is 0.88. If we take into account the uncertainties on both the primary energy of the showers and the estimated electric field at the location of the shower axis, the Pearson correlation coefficient is $0.81^{+0.12}_{-0.46}$ at 95\% CL.

\subsection{A fully reconstructed three-fold coincidence}\label{3fold}
\begin{figure*}[!ht]
\begin{center}
\includegraphics[width=16cm]{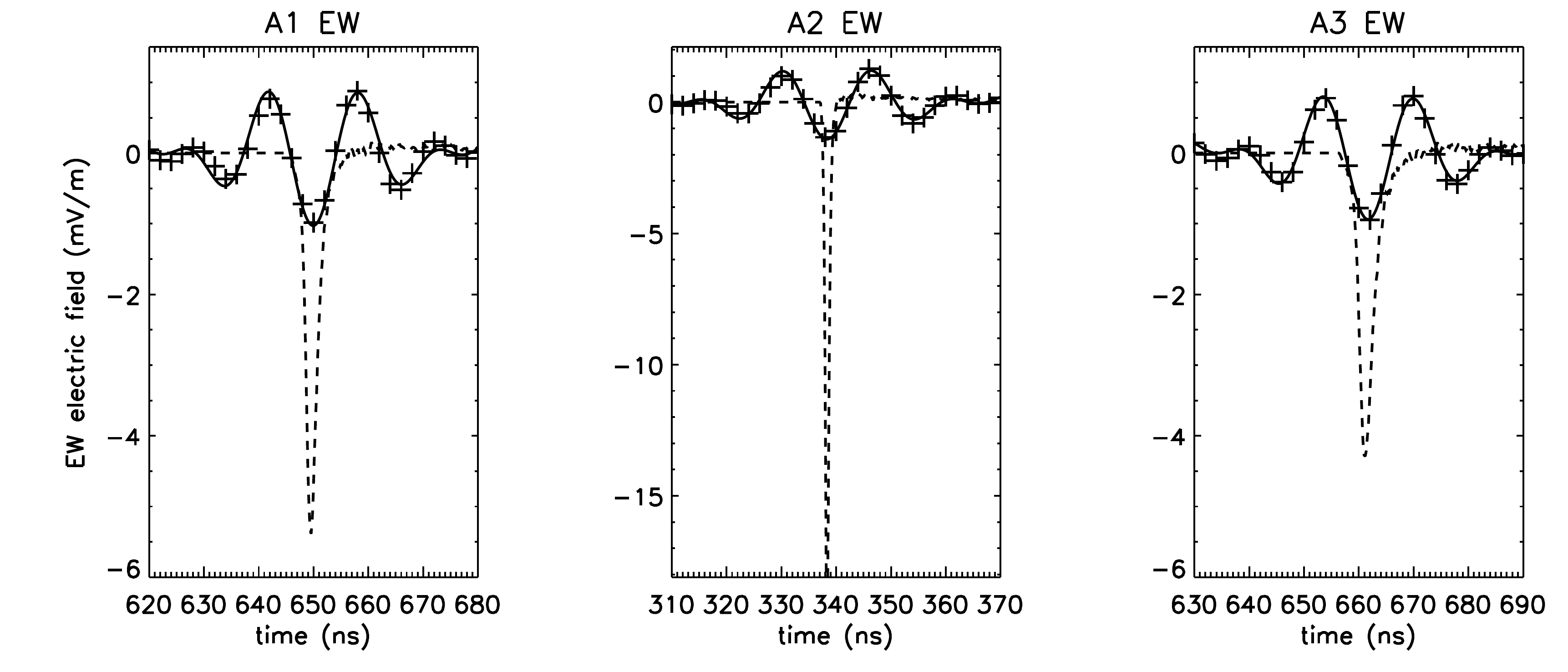}
\caption{\sl{Electric field in the EW polarization for the three radio stations A1, A2 and A3. The dashed line represents the full-band simulated electric field. The solid line is the simulated electric field filtered in the band 40-80 MHz. The data (appearing as crosses) are filtered in the same frequency band and deconvoluted for antenna response. A zoom has been made on the transient region. The simulated electric field amplitude was scaled to match the data (the same normalization factor is used for the three radio stations). The agreement of the pulse shapes between the data and the simulation in the filtered band is very good.}}
\label{sigA1}
\end{center}
\end{figure*}

On November 30, 2009, at 09:45 UTC, a three-fold coincident event of all three antennas with SD was detected.
It was an event detected by five SD stations, including the additional station Apolinario.
The energy of the event is estimated to be $1.4\pm 0.2$~EeV using the standard reconstruction for the SD.
Using the core position and associated errors ($\pm~23$~m and $\pm~47$~m in the EW and NS directions, respectively), the shower axis of the event is at $164\pm 25$~m, $93\pm 18$~m and $188\pm 18$~m from A1, A2 and A3, respectively.
The arrival direction of the shower front of ($\theta=51.3\pm 0.2^\circ$, $\phi=209.8.2\pm 0.1^\circ$) could be determined from the radio-detection setup compared to ($\theta=51.0\pm 0.5^\circ$, $\phi=209.8\pm 0.4^\circ$) for the SD.
The 3D angular difference between the direction estimated by the radio stations and the direction estimated by the SD is
$\delta\alpha=0.4^\circ$, showing that the two directions are perfectly compatible since the SD angular resolution for this type of event is not better than $1^{\circ}$ \cite{angres}.
In Figure~\ref{sigA1}, the data recorded by the three stations for this event are shown.

\begin{figure}[!ht]
\begin{center}
\includegraphics[width=12cm]{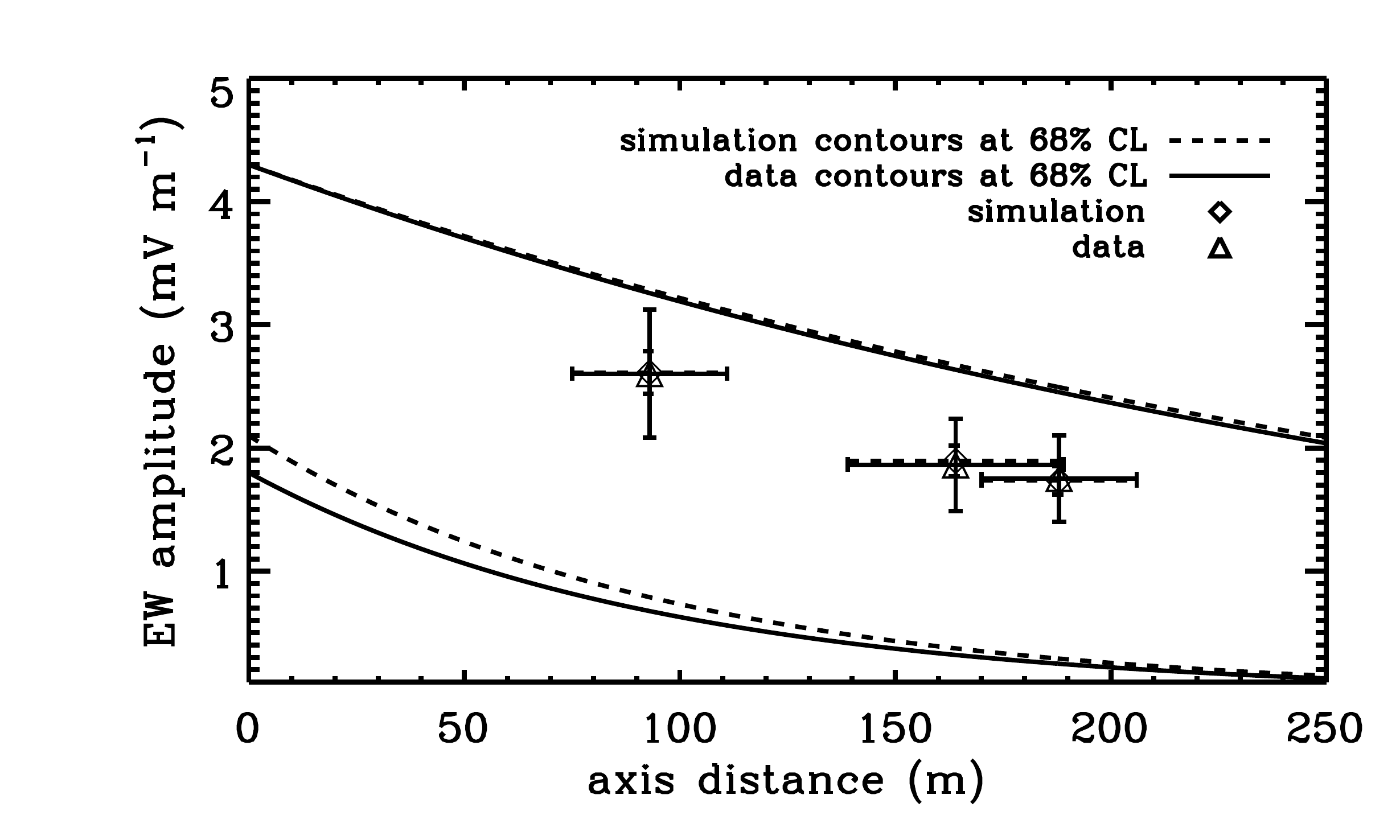}
\caption{\sl{Lateral distribution (profile) of the electric field as a function of the distance to the shower axis for the 3-fold coincident event. The profile is described by an exponential function $\mathcal{E}^{\mathrm{EW}}(d)=\mathcal{E}^{\mathrm{EW}}_0\, e^{-d/d_0}$. The data with error bars are represented by the triangles. The simulated values with error bars are represented by the diamonds. The solid and dashed lines (data and simulation, respectively) represent the profiles using the $\pm 68\%$ $d_0$ values extracted from the Monte Carlo simulations. Electric field values were corrected for the instrumental response.}}
\label{fig:ldf3fold}
\end{center}
\end{figure}

The knowledge of the complete antenna transfer function allows one to convert the voltage obtained by the ADC into the strength of the electric field as received by the antenna (see section~\ref{secant}).
Before deconvolution, the signal was filtered in the band 40-80~MHz, in order to keep clear of large signals from short wave transmitters visible below 35~MHz at this hour of the day.
With the observed deconvoluted signal, fitting an exponential decrease of the electric field amplitude $\mathcal{E}^{\mathrm{EW}}=\max(\mathcal{E}^{\mathrm{EW}}(t))-\min(\mathcal{E}^{\mathrm{EW}}(t))$ with the axis distance $d$ in the EW polarization leads to $\mathcal{E}^{\mathrm{EW}}_0=3.7^{+0.6}_{-1.9}$~mV~m$^{-1}$ at 68\% CL, and $d_0=156^{+175}_{-62}$~m at 68\% CL in the 40-80~MHz frequency band. This profile is presented in Figure~\ref{fig:ldf3fold}.
The errors on both $\mathcal{E}^{\mathrm{EW}}_0$ and $d_0$ were computed by Monte Carlo using varying values of the axis distances according to the SD reconstruction. Concerning the errors on the electric field, we assumed a relative error of 5\% due to the noise conditions (uncorrelated error between the three radio stations), and a relative error of 20\% due to the uncertainty on the soil properties (fully correlated between the three radio stations because they have detected the same event at the same time and with the same soil).

This event has also been simulated with the code SELFAS~\cite{selfas}, using the central values of the event reconstruction parameters (core position, primary energy, arrival direction) and assuming a proton as the primary cosmic ray (with first interaction point $X_1=40~\text{g cm}^{-2}$). A total number of $10^8$ particles in the shower were simulated, so that we can neglect the noise on the final electric field, presented in Figure~\ref{sigA1}. Therefore, a fully correlated error of 20\% (soil conditions) on the three radio stations has been assumed for the computation of the errors on the profile parameters ${\mathcal{E}^{\mathrm{EW}}_{0,\text{SELFAS}}}$ and $d_0^\text{SELFAS}$, in addition to the uncertainties on the axis distances. The same filtered band 40-80 MHz has been used on the simulated data. The exponential profile obtained from the simulation is characterized by
${\mathcal{E}^{\mathrm{EW}}_{0,\text{SELFAS}}=3.2\pm 1.1}$~mV~m$^{-1}$  at 68\% CL
and
$d_0^\text{SELFAS}=175^{+170}_{-80}$~m at 68\% CL, after rescaling of the amplitudes by the same factor on the three radio stations to match the data (see also Figure~\ref{sigA1}).
The large error bars on both $\mathcal{E}^{\mathrm{EW}}_0$ and $d_0$ are mainly due to the relatively small distance between the three radio stations with respect to the axis distances and also on the uncertainties on the axis distances (or equivalently on the shower axis and core position). We see here the limits of having a small array with a weak lever arm.

\begin{figure*}[!ht]
\begin{center}
\includegraphics[scale=0.4]{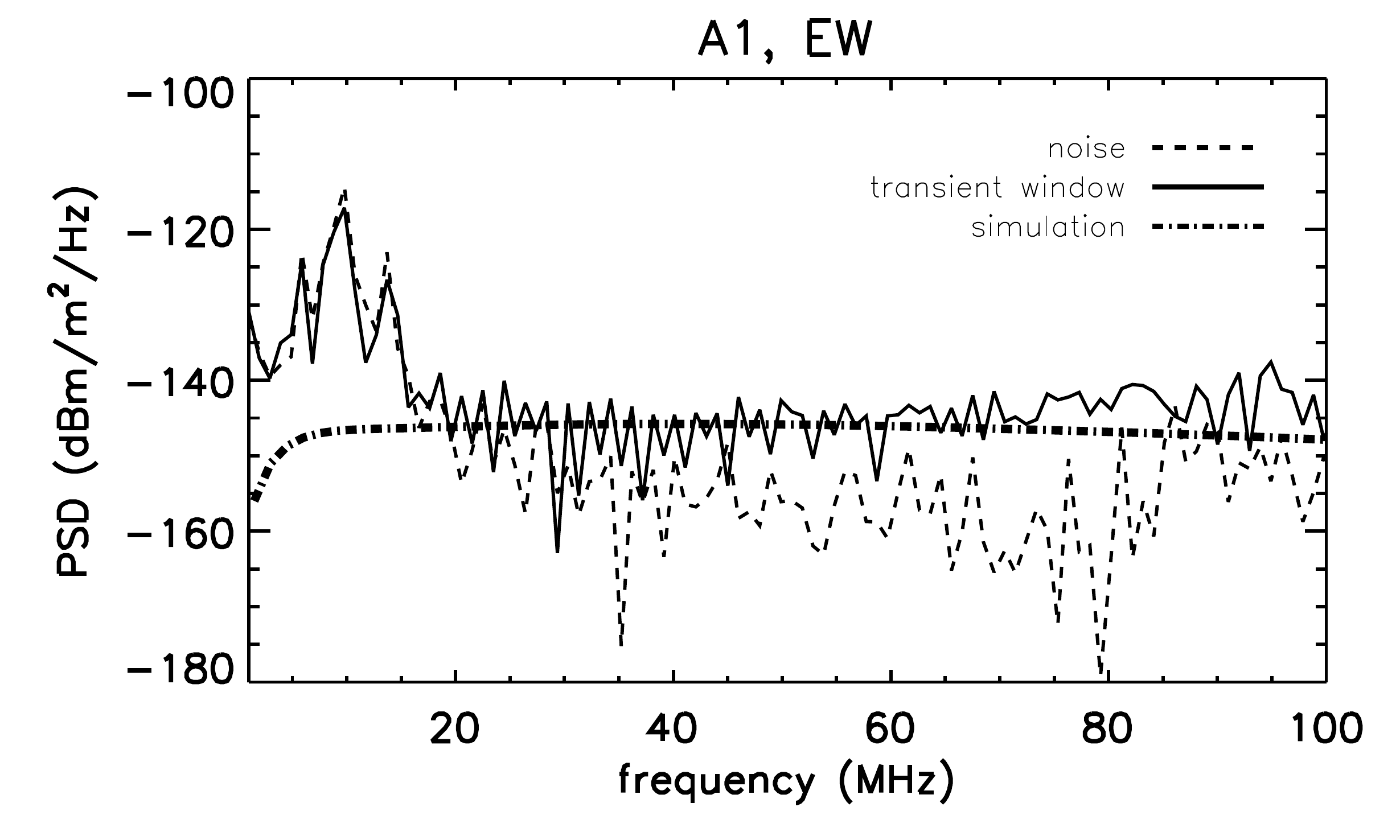}
\includegraphics[scale=0.4]{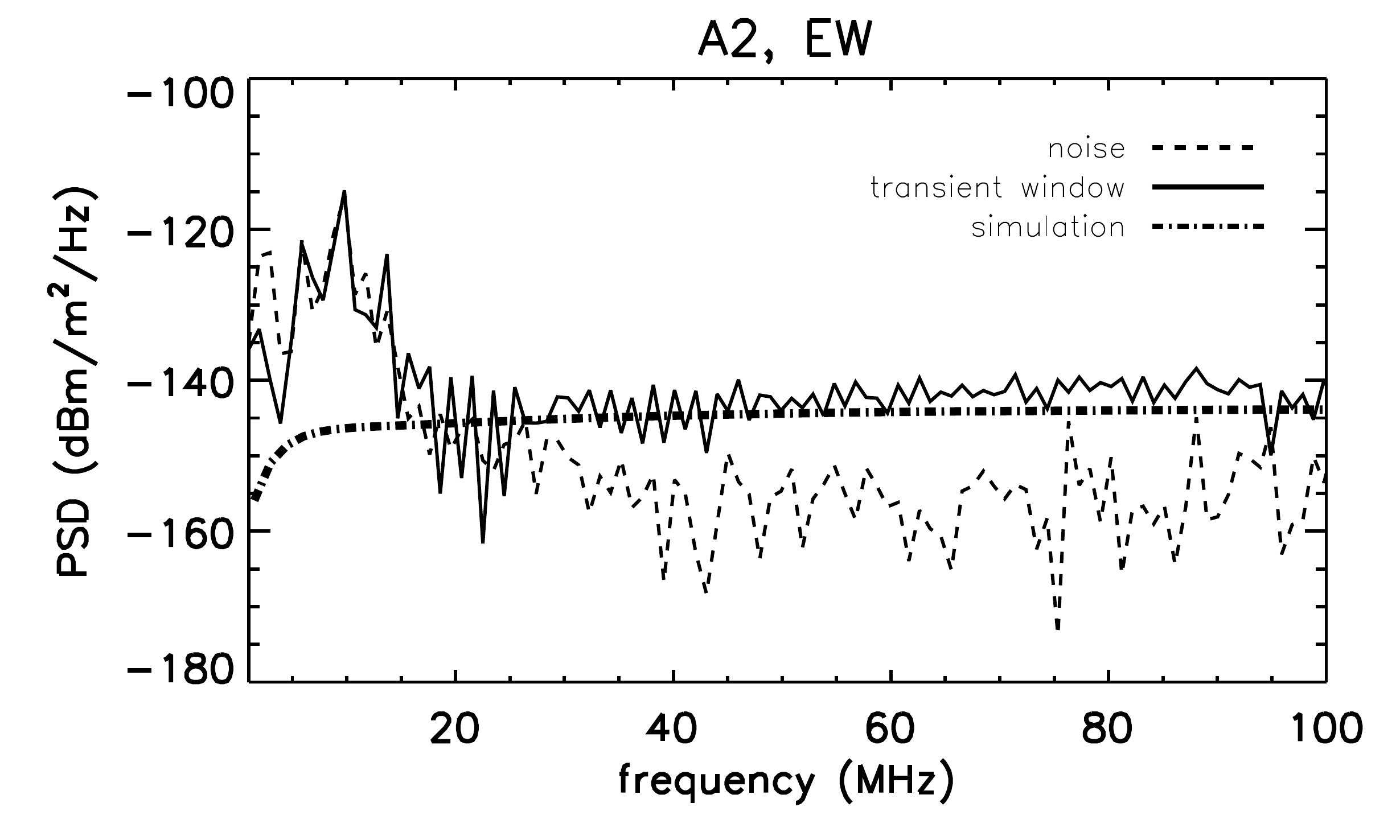}
\includegraphics[scale=0.4]{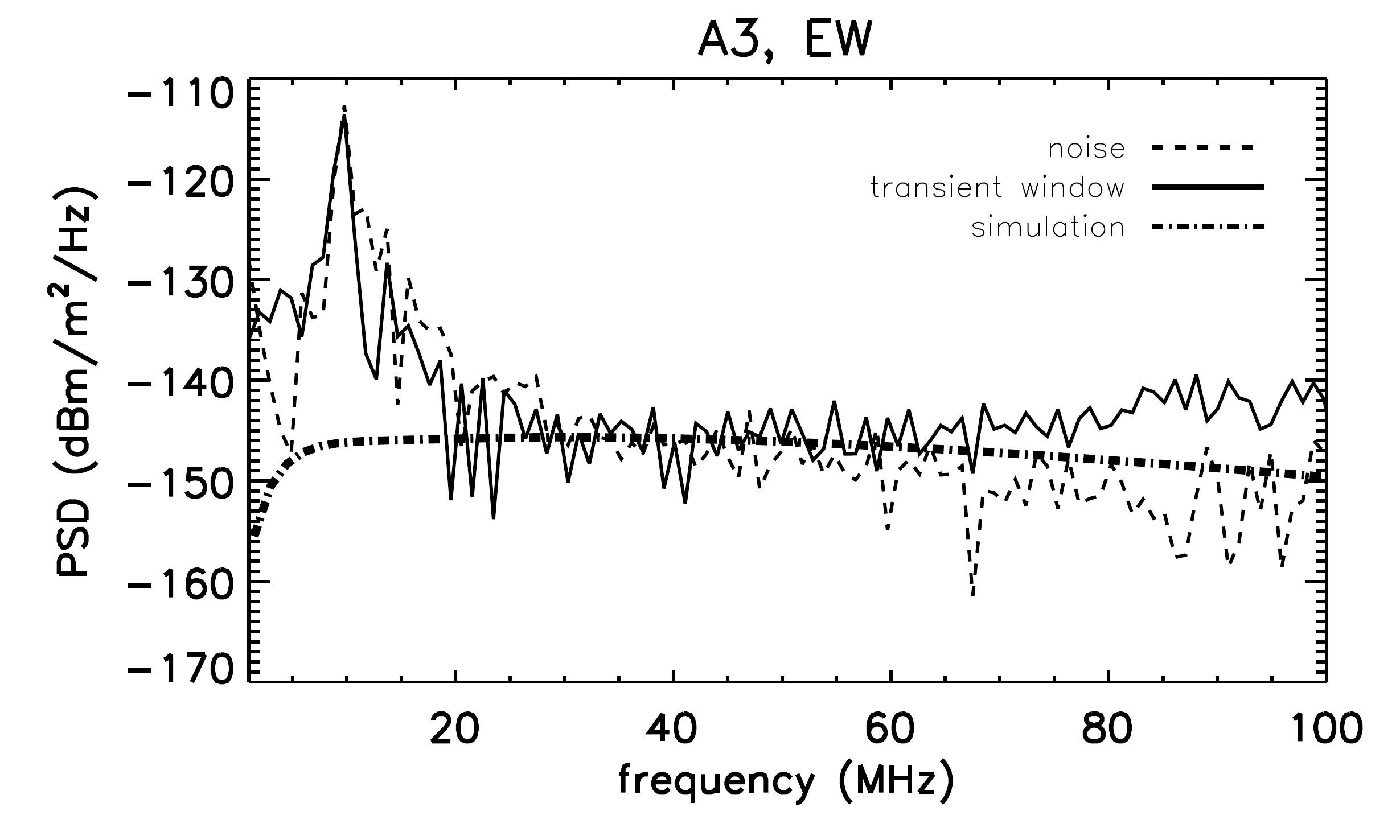}
\caption{\sl{Deconvoluted power spectral density (PSD) of the three radio stations in the EW polarization (top to bottom: A1, A2, A3). The solid line represents the spectrum of the signal in the transient window (512 points around the pulse) and the dashed line corresponds to 512 points outside of the transient window. The spectrum of the simulated electric field is shown as a dot-dashed line. We clearly see the AM emitters below 35~MHz.
No FM signal above 80 MHz is visible at the time of the event.}}
\label{spectra}
\end{center}
\end{figure*}

Finally, Figure~\ref{spectra} shows the deconvoluted power spectral density of the 3 recorded signals in the EW polarization in the noise zone and in the transient zone, both corrected for the system response as in Figure~\ref{sigA1}. The spectra of the transients are computed up to 100~MHz. The spectra of the simulated signals in the EW polarization are superimposed with the data, and show a good agreement when band-limited to 40-80~MHz.

\section{Conclusions}
Between July 2007 and May 2010, the RAuger radio setup installed at the center of the Pierre Auger Observatory has detected, in a fully autonomous and independent way, 65 high-energy cosmic rays.
This first prototype had strong built-in limitations: no dynamic threshold setting and on-site intervention was mandatory to modify the levels; a high deadtime of 2.7~s due to the reading of the trace by the serial link prevented cosmic-ray detection during high event-rate periods;  frequent hardware failures occurred on 2 of them (A2 and A3), and consequently only one three-fold coincident event with the SD was recorded. Nevertheless, RAuger gave valuable results with self-triggered cosmic-ray events. The sky map in local coordinates of the \nevts\ events in coincidence with the SD presents a strong excess of events coming from the south, in agreement with a geomagnetic emission model.
The study of the relative detection efficiency shows that this prototype is particularly sensitive to inclined showers.

The dependence on the electric field as a function of the distance to the shower axis and its
correlation with shower primary energy were studied using a sample of well-reconstructed showers.
The positive correlation is confirmed at a level of 99.99\%.

One three-fold coincidence was detected and its axis direction is fully compatible with the standard reconstruction used for the SD. A test for the dependence of the lateral profile with axis distance has provided the evidence of an exponential decay as first proposed in~\cite{allan}. This event has been compared to the electric field obtained by the simulation code SELFAS, and the agreement is very satisfactory. The recorded pulse shape is compatible with the simulation.

Some further systematic studies were performed and will be investigated in more details in the future.
The influence of the local electric field value on the event rate of such an autonomous station was studied.
The threshold voltage must be high enough at the expense of the lowest energy cosmic-ray detection ability. One of the conclusions of this work is that it will be necessary to use a dynamic threshold for which the efficiency is optimal station by station, especially when the latter cover a large area. In this scope, the next generation of self-triggering stations will use a variable threshold, which will be automatically adjusted as a function of the local background.

\newpage

\section*{Acknowledgments}

The successful installation, commissioning, and operation of the Pierre Auger Observatory
would not have been possible without the strong commitment and effort
from the technical and administrative staff in Malarg\"ue.

We are very grateful to the following agencies and organizations for financial support: 
Comisi\'on Nacional de Energ\'ia At\'omica, 
Fundaci\'on Antorchas,
Gobierno De La Provincia de Mendoza, 
Municipalidad de Malarg\"ue,
NDM Holdings and Valle Las Le\~nas, in gratitude for their continuing
cooperation over land access, Argentina; 
the Australian Research Council;
Conselho Nacional de Desenvolvimento Cient\'ifico e Tecnol\'ogico (CNPq),
Financiadora de Estudos e Projetos (FINEP),
Funda\c{c}\~ao de Amparo \`a Pesquisa do Estado de Rio de Janeiro (FAPERJ),
Funda\c{c}\~ao de Amparo \`a Pesquisa do Estado de S\~ao Paulo (FAPESP),
Minist\'erio de Ci\^{e}ncia e Tecnologia (MCT), Brazil;
AVCR AV0Z10100502 and AV0Z10100522, GAAV KJB100100904, MSMT-CR LA08016,
LG11044, MEB111003, MSM0021620859, LA08015, TACR TA01010517 and GA UK 119810, Czech Republic;
Centre de Calcul IN2P3/CNRS, 
Centre National de la Recherche Scientifique (CNRS),
Conseil R\'egional Ile-de-France,
D\'epartement  Physique Nucl\'eaire et Corpusculaire (PNC-IN2P3/CNRS),
D\'epartement Sciences de l'Univers (SDU-INSU/CNRS), France;
Bundesministerium f\"ur Bildung und Forschung (BMBF),
Deutsche Forschungsgemeinschaft (DFG),
Finanzministerium Baden-W\"urttemberg,
Helmholtz-Gemeinschaft Deutscher Forschungszentren (HGF),
Ministerium f\"ur Wissenschaft und Forschung, Nordrhein-Westfalen,
Ministerium f\"ur Wissenschaft, Forschung und Kunst, Baden-W\"urttemberg, Germany; 
Istituto Nazionale di Fisica Nucleare (INFN),
Ministero dell'Istruzione, dell'Universit\`a e della Ricerca (MIUR), Italy;
Consejo Nacional de Ciencia y Tecnolog\'ia (CONACYT), Mexico;
Ministerie van Onderwijs, Cultuur en Wetenschap,
Nederlandse Organisatie voor Wetenschappelijk Onderzoek (NWO),
Stichting voor Fundamenteel Onderzoek der Materie (FOM), Netherlands;
Ministry of Science and Higher Education,
Grant Nos. N N202 200239 and N N202 207238, Poland;
Portuguese national funds and FEDER funds within COMPETE - Programa Operacional Factores de Competitividade through 
Funda\c{c}\~ao para a Ci\^{e}ncia e a Tecnologia, Portugal;
Romanian Authority for Scientific Research ANCS, 
CNDI-UEFISCDI partnership projects nr.20/2012 and nr.194/2012, 
project nr.1/ASPERA2/2012 ERA-NET and PN-II-RU-PD-2011-3-0145-17, Romania; 
Ministry for Higher Education, Science, and Technology,
Slovenian Research Agency, Slovenia;
Comunidad de Madrid, 
FEDER funds, 
Ministerio de Ciencia e Innovaci\'on and Consolider-Ingenio 2010 (CPAN),
Xunta de Galicia, Spain;
Science and Technology Facilities Council, United Kingdom;
Department of Energy, Contract Nos. DE-AC02-07CH11359, DE-FR02-04ER41300, DE-FG02-99ER41107,
National Science Foundation, Grant No. 0450696,
The Grainger Foundation USA; 
NAFOSTED, Vietnam;
Marie Curie-IRSES/EPLANET, European Particle Physics Latin American Network, 
European Union 7th Framework Program, Grant No. PIRSES-2009-GA-246806; 
and UNESCO.

\end{document}